\documentclass[onecolumn,amssymb, nobibnotes, aps, prfluids]{revtex4-2}
\usepackage{graphicx}
\usepackage{dcolumn}
\usepackage{bm}
\usepackage{hyperref}
\usepackage{amsmath}					
\usepackage{amssymb}
\usepackage{color}
\usepackage{graphicx}
\usepackage{rotating}
\usepackage{multirow}

\usepackage{subcaption} 

\begin{document}


\title{Influence of freestream turbulence and porosity on porous disc-generated wakes}
\author{M. Bourhis}
\author{O. R. H. Buxton}%
\affiliation{Department of Aeronautics, Imperial College London SW7 2AZ, United Kingdom
}

\date{\today}

\begin{abstract}

This study aims to evaluate the effect of freestream turbulence (FST) on wakes produced by discs with different porosity. The wakes are exposed to various freestream turbulence “flavours", where turbulence intensity and integral length scale are independently varied. The turbulent wakes are interrogated through hot-wire anemometry from 3 to 15 diameters downstream of the discs. It is found that discs with low porosity behave similarly to a solid body, both in terms of entrainment behaviour and scaling laws for the centreline mean velocity evolution. Far from the discs, the presence of FST reduces both the wake growth rate and entrainment rate, with a clear effect of both turbulence intensity and integral length scale. As porosity increases, these “solid body" FST effects gradually diminish and are reversed above a critical porosity. The entrainment behaviour in disc-generated wakes is significantly influenced by the presence of large scale coherent-structures, which act as a shield between the wake and the surrounding flow, thus impeding mixing. We found that higher porosity, turbulence intensity, or integral length scale weaken the energy content of these structures, thereby limiting their influence on wake development to a shorter distance downstream of the disc. This, in turn, potentially reduces the influence of large-scale engulfment to the overall entrainment mechanism to a shorter distance downstream of the discs as well. For low porosity discs, freestream turbulence intensity initially promotes near wake growth through the suppression of large-scale structures; however, further downstream, the wakes grow faster when the background is non turbulent.

\end{abstract}


\maketitle


\section{Introduction}

Predicting the development of turbulent wakes is crucial for designing optimal wind farm layouts. As wind farm turbines are arranged in rows, those located in the downstream regions operate within the wake of the upstream turbines, consequently experiencing a reduction in the incoming mean flow velocity and heightened levels of turbulence intensity \cite{Vermeer2003}. This has two major consequences on the operation of a wind farm: firstly, it limits the available momentum for downstream turbines, thereby reducing the power output of a cluster of N turbines compared to the performance of N independent turbines \cite{Barthelmie2007,Barthelmie2010,Hansen2012,PorteAgel2013}. For instance, Hansen~\emph{et al.}~\cite{Hansen2012} reported that turbines located in the downstream regions of the Nysted offshore wind farm may experience a power deficit of up to $50\%$ compared to those in the first row. In addition to these power losses, the turbulence added to the flow by the upstream turbines increases the mechanical stresses and fatigue loading on the blades of downstream turbines, thereby leading to a significant reduction in the fatigue lifetime of wind turbines placed in wakes \cite{Thomsen1999,Frandsen2007,Sorensen2008}. Thomsen \& Sørensen~\cite{Thomsen1999} reported an increase in fatigue load for a turbine installed in a wind farm, ranging between 5\% and 15\% higher than in a free-flow case. Their study reveals that the turbulence intensity of the inflow ($TI_{\infty}$) significantly impacts blade loads, as evidenced by an increase in flapwise blade bending moment with rising $TI_{\infty}$. However, the impact of $TI_{\infty}$ is not limited solely to early blade failure. In their review paper on wind turbine failures, Santelo~\emph{et al.}~\cite{Santelo2021} identified that one of the main causes of torque ripple, ultimately resulting in premature mechanical failure of the generator, is the turbulence intensity of the inflow. Conclusively, this results in heightened maintenance operations for downstream turbines, impacting the overall efficiency of a wind farm. Therefore, to optimise both the layout of a wind farm—number and positioning of turbines within a designated area and wind conditions—and its operation—maximising overall power production whilst minimising maintenance costs—a comprehensive understanding of wake spreading and recovery is essential. 

Historically, studies on wake development have focused on turbulent far wakes generated by solid bluff bodies such as spheres, discs and cylinders placed within a non turbulent background \citep[\emph{e.g.}][]{Hwang1966,Tennekes1972,Cantwell1983,George1989}. However, as wind turbines operate within the atmospheric boundary layer, their wakes do not expand in a quiescent background but instead are exposed to a complex turbulent ambient characterised by its turbulence intensity ($TI_{\infty}$) and integral length scale ($L_{0}$). The wake spreads and grows by mixing with ambient fluid through the process of turbulent entrainment, involving the transport of mass, momentum and energy across the interface between the wake and the neighbouring fluid \citep[\emph{e.g.}][]{Silva2014}. How ambient fluid is “entrained" into the wake affects key characteristics of wake development such as wake width, wake growth rate, evolution of streamwise velocity deficit, mass-flux rate, \emph{etc.}, and introducing turbulence in the ambient flow further complicates the entrainment process. 

In their early review paper on the effect of freestream turbulence (FST) on solid bluff body flows, Bearman \& Morel~\cite{Bearman1983} noted that the presence of FST amplifies the mixing and entrainment processes, thereby increasing the initial spreading rate of the wake. More recently, in two distinct studies, Kankanwadi \& Buxton~\cite{Kankanwadi2020,Kankanwadi2023} investigated the near planar wakes ($x/D \leq 10 $) \citep[in][]{Kankanwadi2023} and far planar wakes ($x/D \approx 40 $) \citep[in][]{Kankanwadi2020} of a circular cylinder subjected to different turbulent “flavours", \emph{i.e.} different $\{TI_{\infty},L_0 \}$ conditions. In particular, they emphasised the distinct roles of $TI_{\infty}$ and $L_{0}$ in determining the physical processes that drive entrainment, \emph{e.g.} engulfment and nibbling. In the near wake region, where the large-scale coherent structures (the von Kármán vortices) remain coherent and energetic, the engulfment of ambient fluid into the wake due to the motion of these large-scale structures prevails over small-scale nibbling. In this region, both the integral length scale and turbulence intensity of the background promote mixing and entrainment, thereby resulting in an increase in the near wake width compared to a non-turbulent ambient, with the integral length scale being the dominant parameter \cite{Kankanwadi2023}. However, in the far wake, Kankanwadi \& Buxton~\cite{Kankanwadi2020} reported a clear reduction in the mean entrainment mass-flux with increasing background turbulence intensity. As the energy content and the coherence of the large-scale structures progressively decays in the wake, the dominance of large-scale engulfment over small-scale nibbling in the entrainment process also diminishes. Similarly, the importance of the integral length scale diminishes, while the intensity of the background turbulence becomes the dominant parameter affecting entrainment behaviour. Consequently, the authors postulate that freestream turbulence intensity has the effect of suppressing, or at least reducing, the entrainment rate in wakes dominated by the nibbling entrainment process. Furthermore, in a control experiment where large-scale coherent vortices were artificially suppressed in the near wake by adding a splitter plate to the rear face of a circular cylinder, Kankanwadi \& Buxton~\cite{Kankanwadi2023} observed a decreased wake growth rate for wakes exposed to FST relative to a quiescent background. Therefore, the prevalence of one entrainment mechanism over the other, along with the impact of FST on the wake development, appears to be primarily determined by the presence and strength of large eddies in the wake, rather than by the physical location within the wake (often denoted arbitrarily as the near and far wake). In summary, Kankanwadi \& Buxton postulate that when the entrainment process is driven by small-scale nibbling (\emph{e.g.} in the ``far wake" of a solid bluff body), FST attenuates entrainment, with $TI_{\infty}$ being the dominant parameter. Conversely, when entrainment is driven by large-scale structures (\emph{e.g.} in the ``near wake" of a solid bluff body), FST enhances entrainment, with the integral length scale $L_0$ becoming the dominant parameter. In a similar vein, Chen \& Buxton~\cite{Chen2023} reported the presence of a turning point in the wake of a cylinder, occurring approximately 15 diameters downstream, wherein the slope of the wake half-width streamwise evolution decreases suddenly. This phenomenon was consistently observed across all FST conditions investigated in this study. The authors hypothesised that this turning point, whose location in the wake likely depends on the dynamics of the near-wake coherent vortices and the Reynolds number, could mark the transition in the entrainment and wake’s growth process from being driven by large-scale engulfment to being driven by small-scale nibbling. Building upon the preceding discussion, it is conceivable that the influence of FST on both the entrainment rate and wake growth rate may vary on either side of this turning point.

While studies on solid bluff bodies provide valuable insights into understanding the physics underpinning wake development in a turbulent background, the applicability of these findings to wind turbine wakes remains questionable. Firstly, the rotor of a wind turbine is not a solid bluff body but can more effectively be modelled as a porous object from a momentum deficit perspective. For instance, Biswas \& Buxton~\cite{Biswas2024} derived an effective porosity based on the blade's geometry and the operating point of the turbine, \emph{i.e.} the tip-speed ratio (ratio of the blade tip velocity to the freestream wind velocity). Moreover, the classical wind turbine aerodynamics theory, known as the General Momentum Theory, relies on the concept of the actuator disc, wherein the rotor is replaced by a permeable thin disc (see \citep[\emph{e.g.}][]{Sorensen2015} for further developments on the General Momentum Theory for wind turbines). In contrast to a solid bluff body, where the drag or thrust coefficient is primarily fixed by its geometry and the inflow Reynolds number (\emph{e.g.} $C_T \approx 1.1 - 1.2$ for a smooth circular cylinder and $C_T \approx 1.1 - 1.3$ for a solid disc at $Re \approx 4 \times 10^5$ \citep[\emph{e.g.}][]{Blackmore2014,Jin2021}), it is possible to adjust the porosity of a porous bluff body to match a thrust coefficient relevant to a wind turbine operating point. As wind turbines operate across a wide range of tip-speed ratios and, consequently, a wide range of thrust coefficients, porous discs with porosity specifically tailored to match the turbine's thrust coefficient stand out as a more faithful substitute for a wind turbine than a solid bluff body. Therefore, in recent decades, there has been a growing interest in using porous discs to simulate wind turbine wakes, owing to their simple structure and ease of setup in wind tunnels, particularly beneficial for studying wind farms layouts with hundreds of models \cite{Camp2016,Camp2019,Charmanski2014,Bossuyt2016,Bossuyt2017}. Porous discs are typically designed to match the diameter and thrust coefficient of typical wind turbine models. Even though no standards exist to date in terms of design, two main types of discs have been used in wind tunnel tests: discs with uniform porosity, sometimes directly cut from industrially manufactured wire mesh \cite{Cannon1993,Aubrun2013,Lignarolo2014,Lignarolo2016,Aubrun2019,Vinnes2022,Vinnes2023}, and non-uniform discs with radially increasing porosity, intended to replicate the decreasing solidity of wind turbine models \cite{Camp2016,Camp2019,Aubrun2019,Helvig2021,Neunaber2021,Vinnes2022,Vinnes2023,Ozturk2023,Kurelek2023}. 

The initial question that arises when employing these wake-generating objects concerns their ability to replicate faithfully a wind turbine wake, a topic extensively explored in the literature \citep[\emph{e.g.}][]{Aubrun2013,Lignarolo2016,Helvig2021,Neunaber2021,Vinnes2022,Ozturk2023,Kurelek2023}. Overall, these studies highlight that porous discs are a good alternative for producing a realistic wind turbine's far wake, despite some differences in the turbulence mixing process \cite{Lignarolo2016} and instantaneous flow phenomena \cite{Helvig2021}. One-point and two point statistical quantities such as the mean velocity deficit, turbulence intensity, integral length scale, mean kinetic energy transport, skewness, and kurtosis of velocity fluctuations have been found to be in reasonable agreement \citep[\emph{e.g.}][]{Aubrun2013,Lignarolo2016,Camp2016,Neunaber2021,Vinnes2022}. Furthermore, some of the higher-order two-point statistical turbulence characteristics of a wind turbine wake, such as the flow intermittency at the centreline and the intermittency ring at the edge of the wake, can be replicated by porous discs \citep[\emph{e.g.}][]{Neunaber2021,Vinnes2023}. 

The general purpose of many of these aforementioned studies was to highlight the similarities and differences between wakes of porous discs and wind turbines. Even though some of these studies were performed in atmospheric-like conditions \citep[\emph{e.g.}][]{Aubrun2013,Neunaber2021} most of them were performed in a non turbulent background as the aim was not so much to quantify the influence of the ambient turbulence “flavours" on the characteristics of the wake, but rather to compare the wake of these two types of wake-generating objects under more realistic inflow conditions. Hence, the impact of freestream turbulence on porous discs wakes has been barely explored in the literature, yet it holds significant interest. This is especially critical for assessing the resemblance in wake properties between porous discs and wind turbines across different “flavours" of freestream turbulence. Moreover, from a more fundamental perspective, it is intriguing to ascertain whether the effects of freestream turbulence on porous discs wakes mirror those observed with solid bluff bodies.

Among the few studies that have investigated the influence of freestream turbulence on wakes generated by porous discs, Vinnes~\emph{et al.}~\cite{Vinnes2023} compared the wakes from 3 diameters to 30 diameters downstream of two distinct porous discs—uniform and non-uniform— and under two different turbulent inflow conditions ($TI_{\infty} = 0.4\%$ and $TI_{\infty} = 4\%$) using hot wire measurements. For both discs, the wake growth rate in the far wake ($x/D \gtrsim 10$) is significantly larger under the low-$TI_{\infty}$ condition compared to the high-$TI_{\infty}$ condition. This observation is in contradiction with the majority of analytical wind turbine wake models, which typically assume an increase in the wake growth rate with increasing turbulence intensity \cite{Bastankhah2014,Niayifar2016}. For the non-uniform disc, the initial wake width is larger in the high-$TI_{\infty}$ case, but due to the lower spreading rate, it becomes significantly larger in the low-$TI_{\infty}$ case for $x/D \geq 5$. For the uniform porous disc, the point at which the wake width is larger in the low-$TI_{\infty}$ is reached further downstream than 30 diameters. Interestingly, similar to what Chen \& Buxton~\cite{Chen2023} reported with a solid bluff body, a turning point can be observed where the slope of the wake half-width streamwise evolution changes suddenly (around 5 diameters downstream for the non-uniform disc). However, the authors observed not a decrease in the wake growth rate, but an increase. Öztürk~\emph{et al.}~\cite{Ozturk2022,Ozturk2023} also examined the near wake characteristics of a non-uniform porous disc, with a geometry closely resembling that used by Vinnes~\emph{et al.}~\cite{Vinnes2023}, under two levels of incoming turbulence intensity ($TI_{\infty} = 0.5\%$ and $TI_{\infty} = 4.5\%$). In contrast to Vinnes~\emph{et al.}, they observed that both the wake width and wake growth rate increase with increasing $TI_{\infty}$; however, one can note that the field of view was limited to 7 diameters downstream of the porous disc. 

While the aforementioned studies offer valuable insights into understanding the impact of FST on porous body-generated wakes, several questions that need to be addressed arise. Firstly, most research investigating the effects of FST on wake development has been conducted with solid bluff bodies, thus focusing on relatively fixed and high thrust coefficients, exceeding those typically achieved by wind turbines. However, it remains uncertain whether the effects of FST are identical for porous bodies achieving lower thrust coefficients and whether these effects are consistent across a wide range of porosity and thrust coefficients. Indeed, in a quiescent background, it has been demonstrated that the flow in the near wake of porous bodies is significantly influenced by porosity \cite{Castro1971,Cannon1991,Xiao2013}, just like the thrust coefficient of a wind turbine strongly affects the initial near wake \cite{Sorensen2015}. For a porous body, the presence/absence and the strength of large-scale coherent structures in the wake, \emph{i.e.} the vortex shedding phenomena, as well as the position and spatial extension of the recirculation region, the initial wake width, velocity deficit, and turbulence generation mechanisms, strongly depends on the porosity. Therefore, it is likely that the effects of FST on the wakes of porous bodies are related to their porosity. Finally, the few studies that have investigated the effects of FST with porous bodies have examined a limited range of porosities and FST conditions, often with turbulence intensities lower than those encountered by wind turbines, and without decoupling the integral turbulent length scale from the turbulence intensity.

Hence, this study aims to fill this gap by conducting an in-depth investigation into the effect of freestream turbulence on the wake of porous discs with varying porosity. Four radially non-uniform porous discs and one solid disc, serving as a reference, were exposed to five different “flavours" of background turbulence $\left\{ TI_{\infty}, L_0\right\}$, where the intensity and integral length scale were varied independently. For all configurations, high-resolution hot-wire anemometry measurements were conducted from 3 to 15 diameters downstream of the discs. 

The paper is structured as follows. The geometries of the discs, the freestream turbulence conditions, and the experimental methodology used to collect the data are outlined in \S~\ref{section:experimental methodology}. The results are subsequently presented and discussed in two distinct sections. \S~\ref{section: characterisation of the wakes} focuses on the effects of disc porosity/thrust coefficient and background turbulence “flavours" on the bulk and entrainment characteristics of disc-generated wakes. An analysis of the flow structures present within the different wakes is also conducted. In \S~\ref{section: wake modelling}, we will investigate the applicability of classical wake models to disc-generated wakes and assess the effects of FST and thrust coefficient on these models.


\section{Experimental methodology \label{section:experimental methodology}}

\subsection{Discs models}

\begin{figure}[b]
    \centering
    \includegraphics[width=0.9\textwidth]{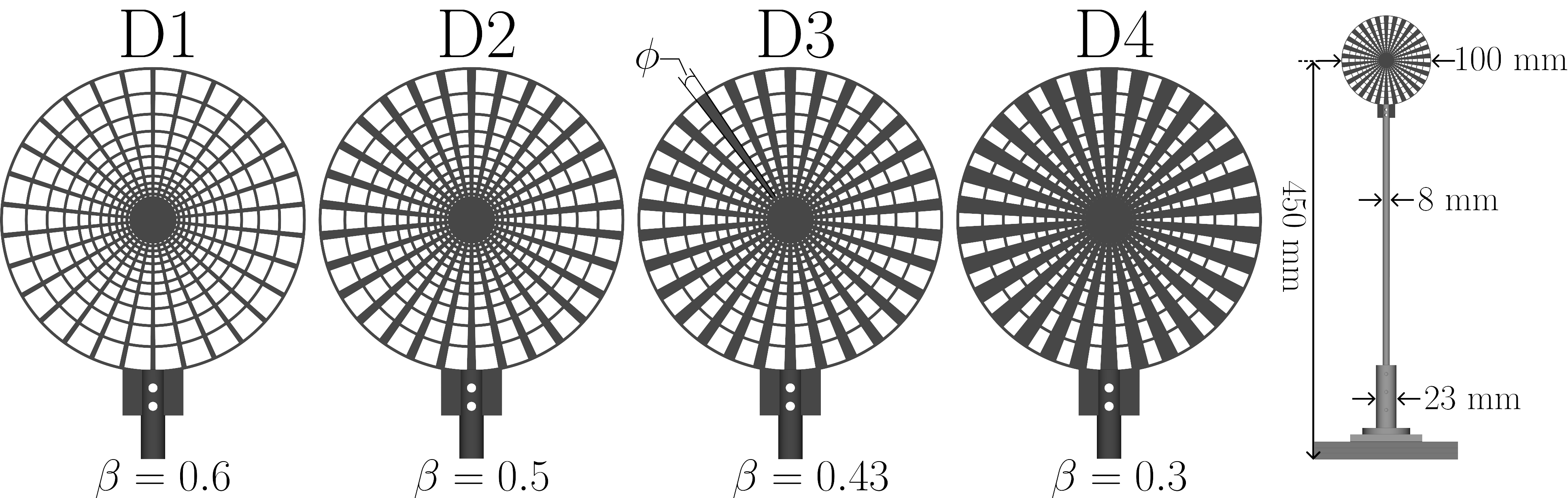}
    \caption{Schematics of the four porous discs}
    \label{FigA}
\end{figure}

Four $D = 100~\text{mm}$ porous discs, each with different porosities—$\beta= 0.6$ for disc 1 (D1), $\beta= 0.5$ for disc 2 (D2), $\beta= 0.43$ for disc 3 (D3), and $\beta= 0.3$ for disc 4 (D4) — and a solid disc (SD) were laser-cut from a 4~mm thick acrylic sheet (FIG.~\ref{FigA}). The porosity $\beta$, defined as the ratio between the open area and the total area of the disc, is varied across each disc by changing the angle $\phi$ formed by the solid spokes extending from the solid centre to the edge of the disc. The design of these porous discs is inspired by the radially non-uniform porous disc initially developed and studied by Camp \& Cal \cite{Camp2016,Camp2019}, which has also been employed in the round robin test organised by Aubrun~\emph{et al.} \cite{Aubrun2019}, and in recent studies by Vinnes~\emph{et al.} \cite{Vinnes2022,Vinnes2023} and Öztürk~\emph{et al.} \cite{Ozturk2022,Ozturk2023}. Camp \& Cal designed a disc that is circumferentially symmetric, with porosity varying radially, to mimic the geometry of a three-bladed rotor used in their experiments (see \cite{Camp2016} for more details on the wind turbine geometry). To converge towards their specific disc geometry, they iteratively adjust the porosity of the disc to align its induction factor and thrust coefficient with that of the rotor operating at a particular tip-speed ratio. Each disc features a solid centre to emulate the nacelle's presence and mirror the increased solidity at the root of the blades. As we move radially outward, the solidity of the disc $\sigma$ diminishes—meaning the porosity increases ($\sigma = 1-\beta$) —thereby reflecting typical wind turbines blade solidity distribution \citep[\emph{e.g.}][]{Sorensen2015}.

All the discs were fastened onto an 8~mm diameter steel rod, which was then fitted and screwed into a larger hollow cylinder made of PLA reinforced with carbon fibers. The larger cylinder was mounted on a 20 mm high aluminium strut profile bolted to the wind tunnel floor. The length of the rod was adjusted in order to position the discs in the middle of the test section, ensuring that the wake of the discs was not affected by the boundary layer produced either by the wind tunnel, the thicker cylinder, or the rail.

\subsection{Facility and experimental procedure}

\begin{figure}[btp]
    \centering
     \begin{minipage}[t]{0.57\textwidth}
        \includegraphics[width=\textwidth]{Figures/FIG2.pdf}
        \caption{Schematic of the experimental setup. The origin of the coordinate system is positioned at he hub centre location.}
        \label{FigB} 
    \end{minipage}
    \hfill
    \begin{minipage}[t]{0.40\textwidth}
        \centering
        \includegraphics[width=\textwidth]{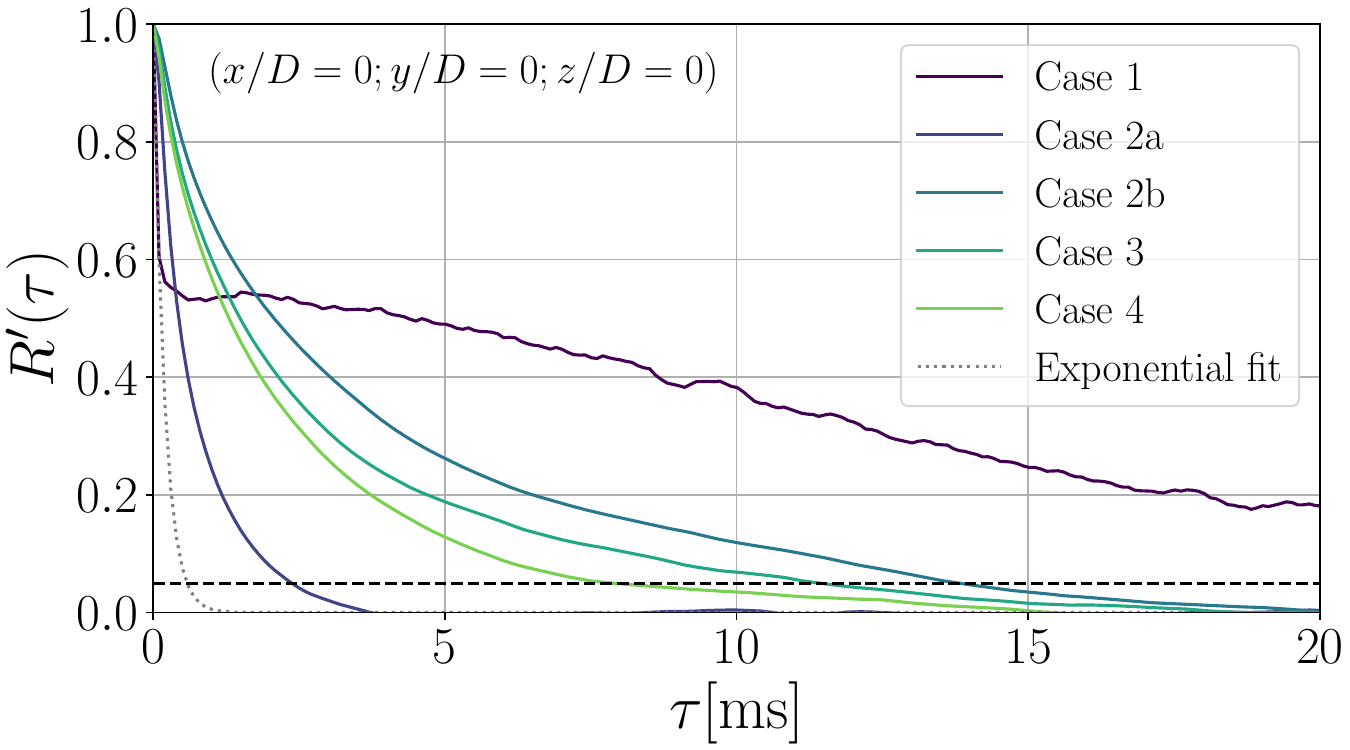}
        \caption{Normalised auto-correlation function of the temporal fluctuations of the streamwise velocity $R'(\tau)$ at the hub centre location for the five FST cases. An exponential function has been fitted to the short-range correlation peak in the “no grid” case 1.}
        \label{FigC}
    \end{minipage}
\end{figure}

The experiments were performed in a closed loop wind tunnel at the Department of Aeronautics, Imperial College London. The tunnel has a test section length of $5~\text{m}$ and a square cross-section of $0.914~\text{m} \times 0.914~\text{m}$. A schematic of the experimental setup, with the coordinate system used in this study, is shown in FIG.~\ref{FigB}. The blockage ratio accounting for the disc, tower, and aluminium rail is around $4\%$.

Velocity measurements were acquired through Hot-Wire Anemometry (HWA) using a Dantec Dynamics 55P11 single-wire probe (5~µm diameter and 1.25 mm long), operated in constant temperature mode by a Dantec 91C10 CTA module and a StreamLine 90N10 frame. The probe was attached to an automated traverse motion system featuring an airfoil-shaped design. The wake was horizontally scanned at the hub height ($z/D=0$) at 6 downstream positions $x/D = 3, 5, 7, 10, 12$, and $15$. Half of the wake was scanned with a measurement spacing of 5~mm ($D/20$), while the other half was scanned with a spacing of 10~mm ($D/10$). The sampling times were set to $60$~s for downstream positions $x/D = 5, 10, 15$, and $30$~s for $x/D = 3, 7, 12$. For every data point, measurements were taken at a sampling rate of 30~kHz and subsequently low-pass filtered using an analog low-pass filter set at 10~kHz. The voltage signal was converted to a velocity signal using a fourth-order polynomial and temperature-corrected according to the methodology proposed by Hultmark and Smits \cite{Hultmark2010}. In this procedure, velocity is no longer solely a function of the hot-wire voltage but also depends on the viscosity and thermal conductivity of the fluid. Hence, the temperature within the wind tunnel and the atmospheric pressure were continuously monitored during the acquisitions using a Furness Control FCO560 micro-manometer. It is worth noting that the temperature rise within the wind tunnel never surpasses $2^\circ \text{C}$ over the course of a day's acquisition. Calibration coefficients were determined using a Pitot probe at the beginning, middle, and end of each campaign day and then temperature-corrected following the aforementioned procedure of Hultmark and Smits. The data were subsequently weighted based on their respective acquisition time within the day. For each velocity profile, the two first and two last measurements outside the wake region were used as the normalising measurements to account for hot-wire drift.

Two turbulence-generating grids of uniform square mesh were placed at different locations upstream of the discs to generate five different “flavours” of incoming freestream turbulence $\{TI_{\infty},L_0\}$. The first grid is a bi-planar wooden grid with a mesh length of 91~mm and a solidity of 36\%. The second grid is a bi-planar acrylic grid with a mesh length of 26.08~mm and a solidity of 22\%. Before conducting experiments with the disc mounted in the tunnel, the flow downstream of the two grids was fully characterised. The discs were then placed at specific locations from the grids to investigate the parameter space $\{TI_{\infty},L_0\}$ as widely as possible.

In the following, we use the Reynolds decomposition to separate the velocity into a time-averaged and a fluctuating component: $u(t) = \langle u(t) \rangle + u'(t) = U + u'(t)$. The turbulence intensity $TI$ is defined as the ratio between the standard deviation of the velocity fluctuations and the mean velocity :  $TI = \sqrt{ \langle u'^{2} \rangle }/U$. Additionally, we denote with the subscript $\infty$ the characteristics of the freestream, \emph{e.g.} $U_{\infty}$ and $TI_{\infty}$ for the bulk freestream speed and turbulence intensity of the background (see Eq.~\ref{Eq1}). The incoming wind speed was set at $U_{\infty}=10~\text{m.s}^{-1}$, resulting in a Reynolds number based on $D$ and $U_{\infty}$ around $68,500$, a regime at which the velocity deficit and far wake turbulence intensity generated by the discs are independent of the Reynolds number \cite{Chamorro2012}. The integral length scale of the turbulent freestream $L_{0}$ is determined by estimating the integral time scale $T_{0}$ and by applying Taylor’s hypothesis of frozen turbulence, \emph{i.e.} $L_{0} = T_0 U_{\infty}$. $T_0$ is determined by integrating the normalised auto-correlation function of the temporal fluctuations of the streamwise velocity with time lag $\tau$, denoted as $R'(\tau)$, up to $\tau^{\star}=0.05$ (Eqs.~\ref{Eq2} \& \ref{Eq3}). FIG.~\ref{FigC} depicts the normalised auto-correlation function at the hub location for the 5 different FST cases. In Case 1, corresponding to the “no grid" scenario, the velocity correlation persists over a large time lag typical of predominantly laminar flow, thus $T_0$ was computed in the same manner as Kankanwadi \& Buxton \cite{Kankanwadi2020}, by fitting an exponential to the short-range correlation peak. 

\begin{figure}[h]
    \centering
     \begin{minipage}[c]{0.5\textwidth}
        \includegraphics[width=0.85\textwidth]{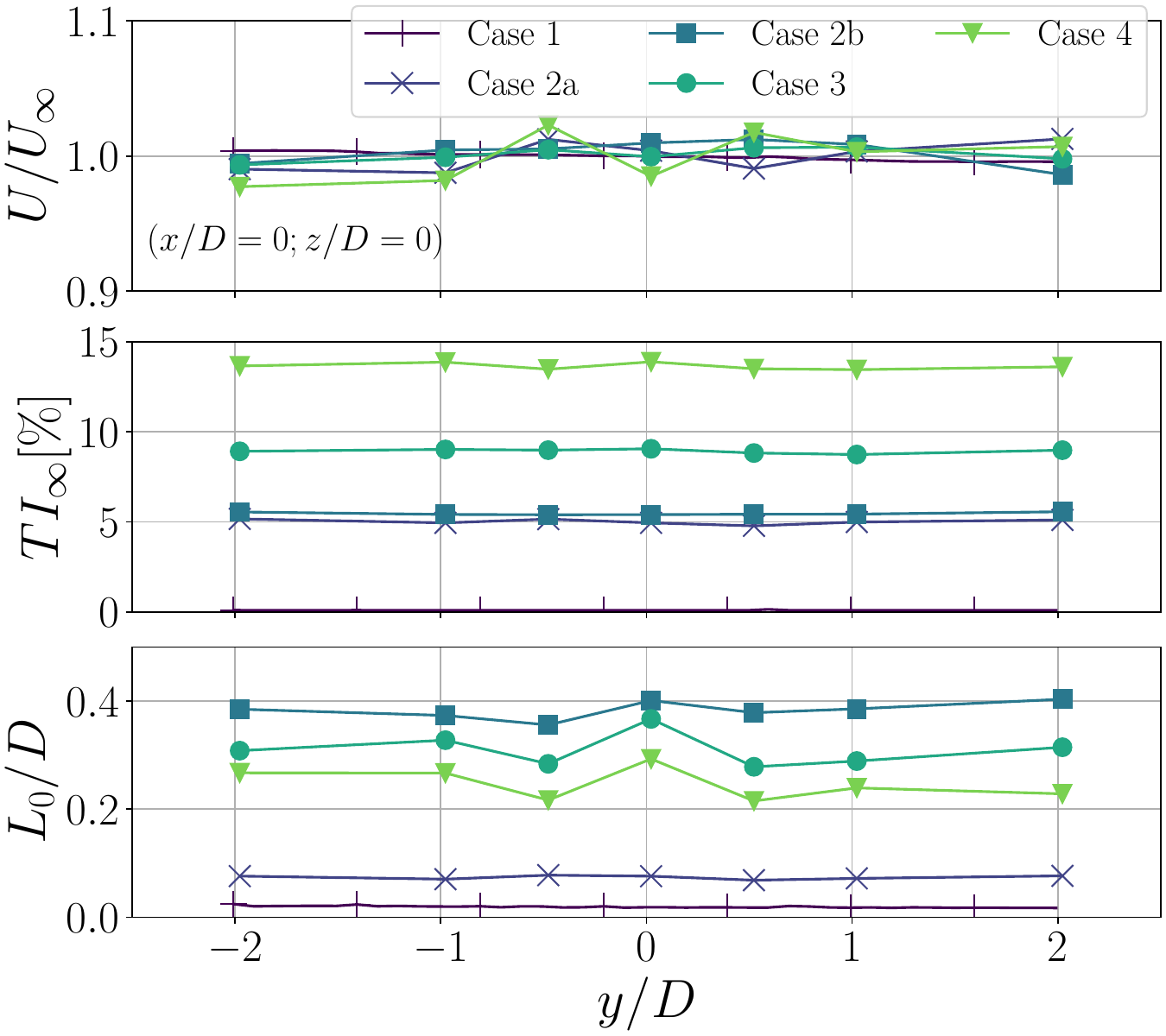}
        \caption{Horizontal profiles of normalised mean velocity $U/U_{\infty}$, freestream turbulence intensity $TI_{\infty}$, and integral length scale $L_0/D$ at the hub centre height and streamwise position ($\{x/D;z/D\} = \{0;0\}$) for the five FST cases.}
        \label{FigD}
    \end{minipage}
    \hfill
    \begin{minipage}[c]{0.45\textwidth}
        \centering
        \includegraphics[width=\textwidth]{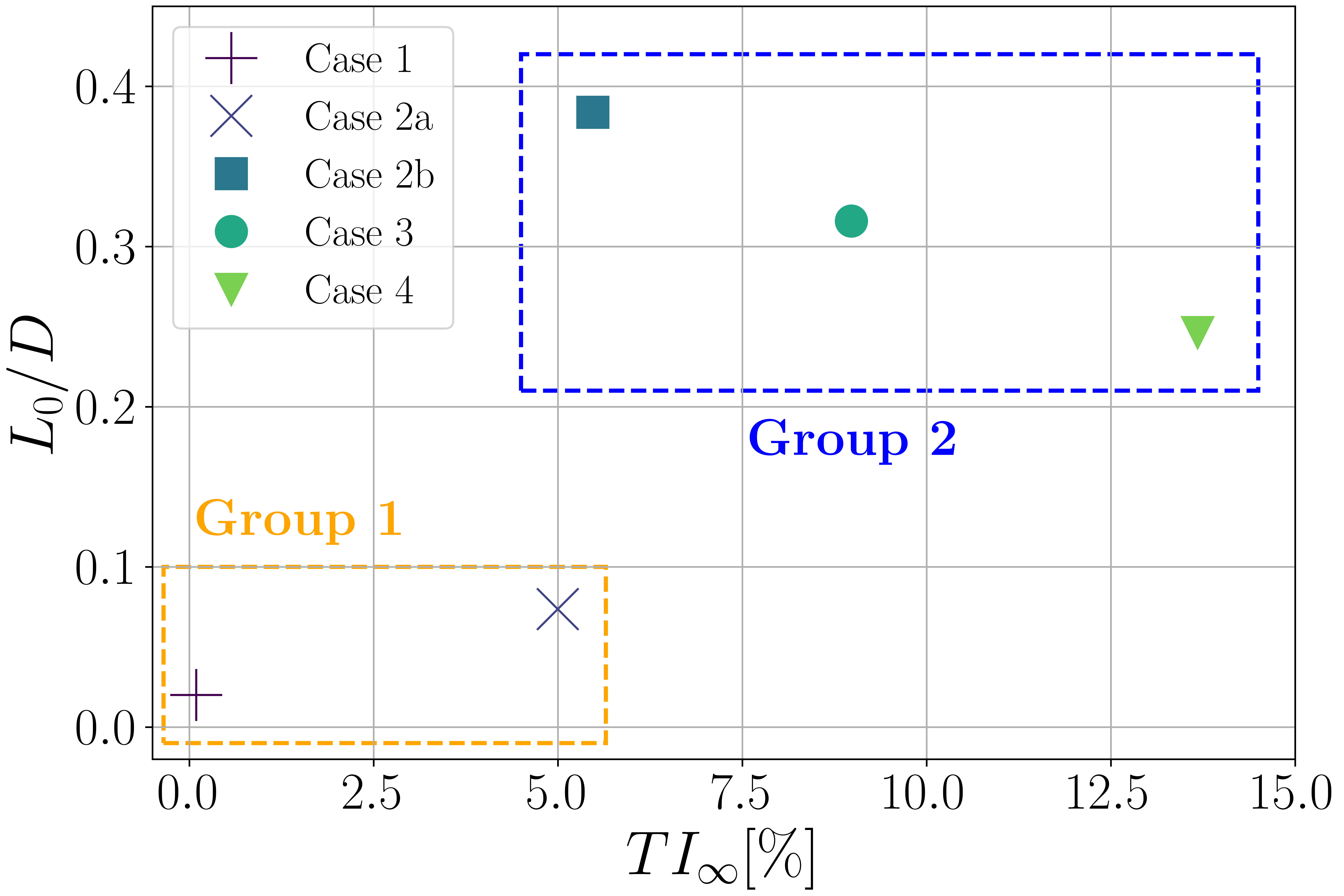}
        \caption{Experimental envelope of freestream turbulence characteristics $\{TI_{\infty},L_0/D\}$ for the five FST cases. \emph{Nb}: Cases 2a and 2b provide a meaningful comparison of the effect of length scale at constant turbulence intensity.}
        \label{FigE}       
    \end{minipage}  
\end{figure}

FIG.~\ref{FigD} shows the horizontal profiles of $U/U_{\infty}$, $TI_{\infty}$ and $L_0/D$ obtained at the hub centre location ($\{x/D;z/D\} = \{0;0\}$) with the tower positioned in the tunnel. The mean velocity, turbulence intensity, and integral length are very uniform. Minor variations in $L_0/D$ are noticeable near $y/D = 0$ presumably because of the tower's presence. The experimental envelopes of the 5 different turbulence “flavours” $\left\{ TI_{\infty},L_{0} \right\}$ shown in FIG.~\ref{FigE} are computed by averaging the horizontal profiles of $TI_{\infty}$ and $L_0$ over the range $-2 \leq y/D \leq 2$. The five FST cases cover a range of freestream turbulence intensities from 0.3\% to 13.7\%, with length scales ranging from 0.03 up to 0.4 disc diameters. Without grids in the wind tunnel, the flow at the disc location is primarily laminar, featuring very low turbulence intensity ($TI_{\infty} \approx 0.3\%$) and integral length scale ($L_0/D \approx 0.03$). Case 2a was achieved using the acrylic grid with a finer mesh, allowing the generation of lower integral length scale, whereas cases 2b, 3, and 4 were obtained with the wooden grid positioned at various upstream distances from the disc. To distinguish the effect of turbulence intensity and length scale, we strategically placed the wooden grid at a specific location (case 2b), where $TI_{\infty}$ at the disc location is identical to that of case 2a but with an integral length scale five times larger. The FST cases have been categorised into two different groups. Group 1 comprises cases 1 and 2a (thin markers), characterised by low integral length scales and relatively low turbulence intensities, while Group 2 comprises cases 2b, 3, and 4 (thick markers), characterised by higher turbulent integral length scales and intensities.

\begin{align}
    TI_{\infty} &= \frac{\sqrt{ \langle u'^{2} \rangle }}{U_{\infty}} \label{Eq1} \\ 
    R'(\tau) &= \frac{\langle  u'(t)u'(t+\tau) \rangle }{\sqrt{\langle u'(t)^2 \rangle}\sqrt{\langle u'(t+\tau)^2 \rangle}}   \label{Eq2} \\
    T_0 &= \int_0^{\tau^{\star}} R'(\tau) d\tau  \label{Eq3} \\   
    L_0 &= U_{\infty} T_0  \label{Eq4}
\end{align}

\begin{figure}[t]
    \begin{minipage}[c]{0.7\textwidth}
    \centering
    \begin{ruledtabular}
    \captionof{table}{Thrust coefficients produced by the discs, calculated from the integration of the momentum deficit at $x/D=7$ in case 1 (non-turbulent background). \label{Tab1}}
    \begin{tabular}{ccc ccc}
     & D1 & D2 & D3 & D4 & SD \\
     $C_T$ & 0.19 & 0.55 & 0.58 & 0.71 & 1.18 
    \end{tabular}
    \end{ruledtabular}
\end{minipage}
\end{figure}

\section{Characterisation of the wakes  \label{section: characterisation of the wakes}}

\subsection{Thrust coefficient}

The thrust coefficients, $C_T = T / (0.5 \rho \pi (D/2)^2 U_{\infty}^2)$, produced by the discs in the non-turbulent background case (Case 1), were estimated using global momentum theory (TABLE~\ref{Tab1}). This method has been widely employed to experimentally estimate the thrust produced by solid discs \cite{Carmody1964}, porous discs \cite{Camp2016,Lingkan2023} and rotors \cite{Aubrun2013}. The thrust $T$ is obtained by integrating the momentum deficit flux across a reference surface :
\begin{align}
    T  =  2 \pi \rho \int U (U_{\infty} - U) r dr \label{EqA}
\end{align}
It is worth noting that the assumption of pressure recovery is implicitly utilized in the computation of thrust. Aubrun~\emph{et al.}~\cite{Aubrun2013} reported that the difference between the static pressure in the wake of a rotor/porous disc ($\beta=0.55$), and the static pressure of the undisturbed free flow, becomes negligible once $x/D \geq 1.5$. Additionally, Carmody~\emph{et al.}~\cite{Carmody1964} experimentally confirmed the validity of Eq.~\ref{EqA} for a solid disc when $x/D \geq 6$. Therefore, for consistency across the discs, we estimated the thrust coefficients produced by the discs by integrating the momentum fluxes at $x/D = 7$ up to the radial coordinate where the velocity deficit is $1\%$ the maximum velocity deficit. The thrust coefficient produced by the solid disc ($C_T \approx 1.18$) closely matches values reported in the literature for laminar flow within a comparable range of Reynolds number, \emph{e.g.} $C_T = 1.14$ \cite{Carmody1964}, $C_T = 1.17$ \cite{Roos1971}, $C_T = 1.17$ \cite{Blackmore2014}. Calculating the momentum deficit appears to be an appropriate approach for estimating a representative thrust coefficient, thus we applied this method to derive $C_T$ for the entire set of discs. The range of $C_T$ produced by these discs mirrors the diverse operating points and thrust coefficients reached by a wind turbine (see \citep[\emph{e.g.}][]{Medici2006,IEA2020} for a typical thrust coefficient \emph{vs.} wind speed curve of lab-scale and full-scale wind turbines). Low thrust coefficients are typically achieved at high wind speeds when the turbine approaches its rated power and is then regulated to convert less wind power, whereas thrust coefficients close to $C_T \approx 0.8$ are commonly linked to operation at the design point and maximum efficiency.

It is worth noting that in the present study, distinguishing between the effects of porosity and thrust coefficient on wake development is impossible, despite both parameters potentially being important in determining the wake dynamics. However, one can note that while there is a fairly similar decrease in porosity between the different porous discs, the difference in thrust coefficient is substantially more pronounced between D1 and D2 compared to the other porous discs. Specifically, D1 produces a considerably lower thrust coefficient despite its porosity not differing significantly, particularly compared to D2. This emphasises the intricate relationship between the thrust and near wake generated by a porous body and its porosity -- a relationship likely dependent on both the overall porosity and its distribution within the geometry. Even in the case of complex porous bodies such as wind turbines, Dong~\emph{et al.}~\cite{Dong2023} observed that rotors with identical thrust coefficient but different blade geometries, and consequently different porosities, generate different wakes. For brevity, in the following analysis, we will refer to the discs by their porosity rather than their thrust coefficient. However, we will discuss further this aspect in the conclusion.

\subsection{Evolution of the velocity deficit and turbulence intensity}

\begin{figure}
    \centering
    \begin{subfigure}{1\textwidth}
    \centering
    \includegraphics[width=\linewidth]{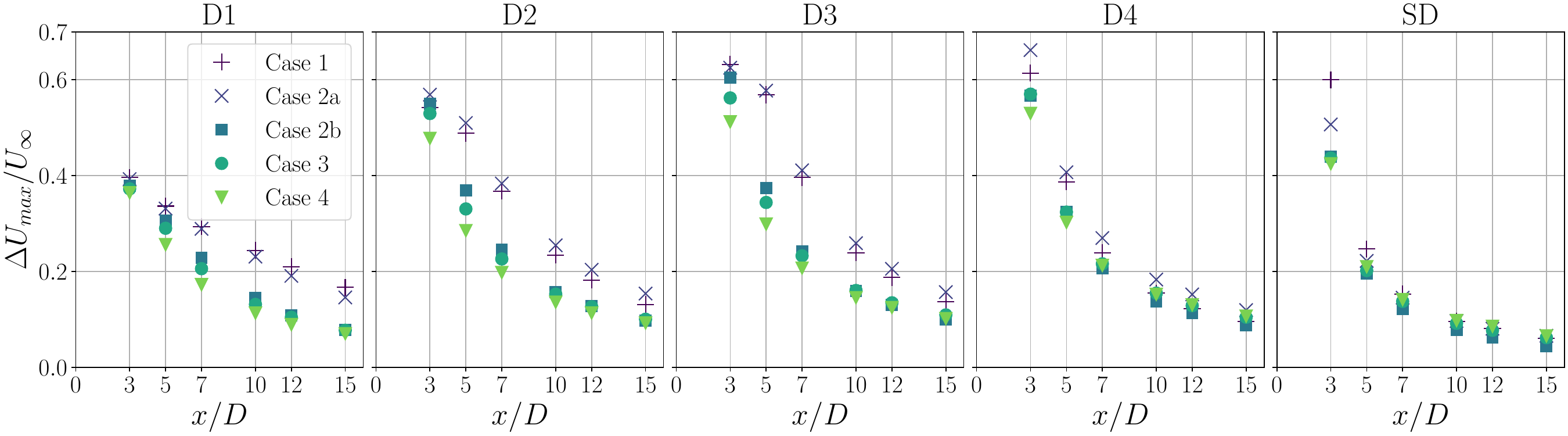}
    \caption{}
    \label{Fig:MaxVeloDef}
  \end{subfigure}
  \begin{subfigure}{1\textwidth}
    \centering
    \includegraphics[width=\linewidth]{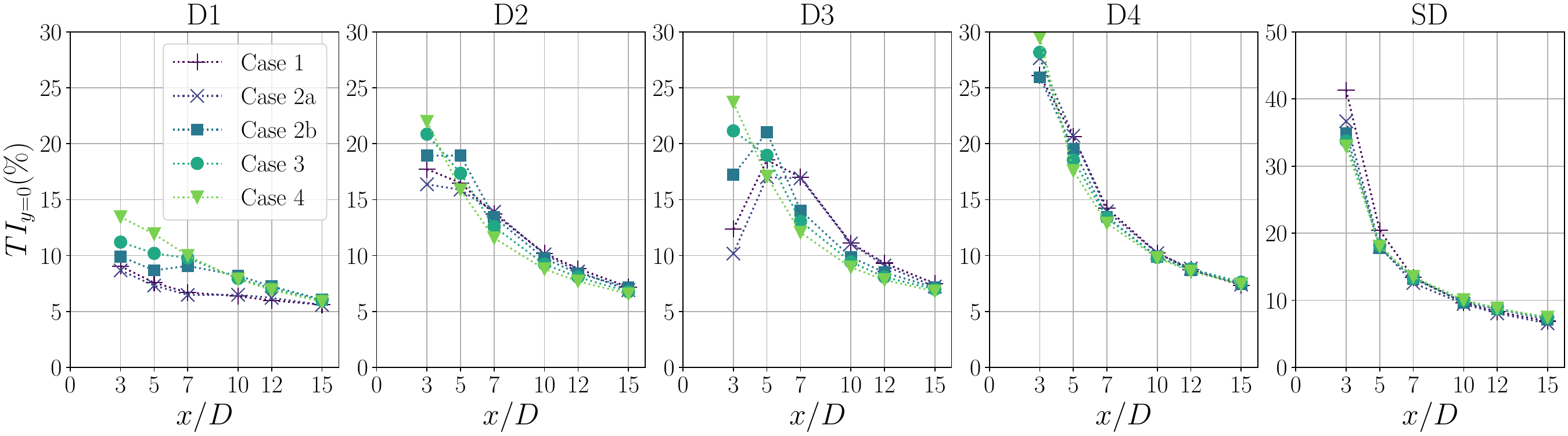}
    \caption{}
    \label{Fig:TIcentreline}
  \end{subfigure}
      \caption{Evolution of (\subref{Fig:MaxVeloDef}) the maximum velocity deficit $\Delta U_{max}/U_{\infty}$ and (\subref{Fig:TIcentreline}) the centreline turbulence intensity $TI_{y=0}$ with increasing downstream distance $x/D$. Note the change in the ordinate axis for the $TI_{y=0}$ graph for SD.}
    \label{Fig:TI_Velodef}
\end{figure}



The normalised mean velocity deficit profiles ($\Delta U/U_{\infty}$) and turbulence intensity profiles ($TI$) at $x/D = 3 - 10 - 15$ in the wakes of all 5 discs, and for the different turbulence “flavours”, are shown in a large scale in Appendix~\ref{AppendixA}. In the near wake, at $x/D =3$, the shape of the velocity deficit profiles varies depending on the thrust coefficient produced by the discs, and the FST conditions. In certain $\{\textrm{disc}/\textrm{FST case}\}$ configurations, the velocity deficit at $x/D=3$ has a top-hat Jensen-like profile (see \emph{e.g.} D3/Cases 1 \& 2a), a rounded but not Gaussian-like profile (see \emph{e.g.} D2/Cases 1 \& 2a), or a Gaussian-like profile (see \emph{e.g.} in Case 4 for all discs). For all FST cases and discs, in the far wake — let us consider, as a first approximation, a distance of 10 diameters downstream of the discs — the velocity deficit profiles exhibit Gaussian-like shapes with magnitudes decaying and widths increasing with distance from the discs. The presence of background turbulence accelerates the transition of the velocity deficit profiles from a top-hat shape or a non-Gaussian shape in the near wake to a Gaussian shape in the far wake, indicating that the turbulence in the wake is developing faster when the ambient flow is turbulent.

The shapes of $TI$ profiles exhibit also some variations depending on the porosity of the discs and the FST conditions. For certain $\{\textrm{disc}/\textrm{FST case}\}$ pairs, the $TI$ profiles have already reached a developed state at $x/D=3$, in a sense that the shapes remain consistent from the near wake to the far wake but with a decreased magnitude and a larger width (see \emph{e.g.} D2, D3 and D4 in Case 4). However, for some other $\{\textrm{disc}/\textrm{FST case}\}$ couples, the $TI$ profiles close to the discs display two significant peaks in the shear layer (see \emph{e.g.} Cases 1 \& 2a for D3 and D4).
It is worth noting that since the discs are axisymmetric, the wakes have a helical structure \cite{Zhang2023} (similarly to a wind turbine tip vortex system), which explains why we observe two peaks of turbulence intensity in the shear layer when interrogating horizontal slices of the wakes. The presence of these peaks suggests that the shear layer emanating from the disc's edge has not yet merged and that the turbulence is still building in the wake. The difference between the magnitude of these peaks and the turbulence intensity at the centreline is amplified at low levels of turbulence in the background, showing that the speed at which the shear layer spreads towards the wake's centreline is increased in the presence of background turbulence. It is likely because of the elevated levels of turbulence intensity produced by the discs in the presence of FST which promote the faster homogenisation of $TI$ within the wake. 

Moreover, when comparing discs D3 and D4 in cases 2a and 2b, the highest turbulence intensity generated by D4 also promotes the early merging of the shear layer in the wake, thereby explaining the lower relative intensity of the peaks in the shear layer compared to D3. For the solid disc, the shear layer has already merged at $x/D=3$ for all FST cases. This is also likely a result of the elevated levels of turbulence intensity produced by this disc, which promote the mixing of the wake, and ultimately leads to the early merging of the shear layer, and a faster homogenisation of the turbulence intensity in the central region of the wake. Ultimately, for all discs, the presence of background turbulence speeds up turbulence evolution in the wake, with the $TI$ profiles reaching a developed state closer to the discs.


FIG.~\ref{Fig:MaxVeloDef} and FIG.~\ref{Fig:TIcentreline} respectively display the streamwise evolution of the maximum velocity deficit $\Delta U_{max}/U_{\infty}$, and the turbulence intensity at the centreline $TI_{y=0}$ for all $\{\textrm{disc}/\textrm{FST case}\}$ pairs. As expected, an increase in flow blockage resulting from decreased porosity leads to a higher initial maximum velocity deficit and a larger turbulence intensity introduced to the flow by the discs. The lower initial maximum velocity deficit observed in FIG.~\ref{Fig:MaxVeloDef} at $x/D = 3$ downstream of the solid disc compared to D4 is likely due to the earlier onset of the mixing process and wake recovery. Moreover, when comparing D2 and D3 in Cases 1, 2a, and 2b, one can note that $TI_{y=0}$ at $x/D = 3$ is larger for D2 than for D3 (FIG.~\ref{Fig:TIcentreline}). However, the turbulence downstream of D3 is still building, and once the shear layer has merged and the turbulence begins to decay, the maximum $TI_{y=0}$ for D3 is consistently higher than that of D2 across all FST conditions.

Both the velocity deficit recovery and turbulence evolution are enhanced in the wake of low porosity discs (D4 \& SD), likely due to the high turbulence production in the near wake. As porosity decreases, the turbulence intensity produced by the discs increases, thereby resulting in a more significant difference between the turbulence intensity in the wake $TI$ and that in the ambient flow $TI_{\infty}$. This, in turn, appears to enhance turbulence mixing and the entrainment of high-momentum, low-turbulent-intensity fluid from the ambient flow into the wake, as illustrated by the faster decay of turbulence intensity and maximum velocity deficit in the near wake of low porosity discs. The additional impact of ambient turbulence in wake recovery is reduced for low porosity discs, and is confined closer to the discs, due to the high level of turbulence generated by them, which enhances near wake mixing. For instance, in the case of the solid disc SD and the disc with the lowest porosity D4, the velocity deficit and $TI$ profiles approximately collapsed in the central wake region for $x/D \geq 10$ across all FST conditions (FIG.~\ref{Fig_AppendixA} \& FIG.~\ref{Fig_AppendixB} in Appendix~\ref{AppendixA}). This demonstrates that, far from these discs, the wake recovery and turbulence evolution are less affected by the ambient conditions and are primarily driven by the turbulence generated by the discs themselves. However, for the disc D1, which generates the lowest initial velocity deficit and turbulence intensity, discrepancies in the central wake region among the different FST cases persist into the far wake ($x/D \geq 10$). Given the low turbulence intensity generated by this disc, the additional impact of FST on wake recovery is more pronounced and persists over a greater distance downstream of the disc. As the wake develops further downstream, turbulence intensity within the wake tends to homogenize with that of the ambient flow. Consequently, a slower decay of centreline turbulence intensity is observed in the far wake for all discs due to a weakened turbulence intensity gradient between the wake region and the ambient flow.

Focusing now on the effects of FST, it is observed that, for all discs, the maximum velocity deficit $\Delta U_{max}/U_{\infty}$ and turbulence intensity at the centreline $TI_{y=0}$ decay faster for Group 2 FST cases, characterised by higher turbulent integral length scales $L_0$ and turbulence intensity $TI_{\infty}$ (Cases 2b, 3 \& 4), compared to Group 1 FST cases (Cases 1 \& 2a). At all six streamwise positions $x/D$, the maximum velocity deficit is lower for Group 1 cases than for Group 2 cases (FIG.~\ref{Fig:MaxVeloDef}). This trend is consistent with the highest levels of turbulence intensity in the near wake in the presence of background turbulence (FIG.~\ref{Fig:TIcentreline}), which promote the mixing of the ambient fluid with the wake, thereby accelerating velocity deficit recovery and turbulence evolution. 

Furthermore, when comparing cases 2a and 2b (which have the same $TI_{\infty}$ but $(L_0)_{\textrm{Case 2b}} \approx 5(L_0)_{\textrm{Case 2a}}$), it is clear that ambient turbulence with a higher integral length scale promotes a faster velocity recovery and a quicker decay of the centreline turbulence intensity. This trend is observed for the whole set of discs, and thus appears to be independent on the thrust coefficient/disc porosity, at least for the range of $L_0$ examined in this study. Given that wind turbines operate in turbulent conditions where the integral length scale can exceed several times their diameter, it would be valuable to extend the range of $L_0$ investigated in future studies.

At high integral length scale (Group 2 FST cases), increasing the intensity of the background turbulence $TI_{\infty}$ accelerates the recovery of velocity deficit, and the decay of the centreline turbulence intensity. It can be observed in FIG.~\ref{Fig:MaxVeloDef} that $(\Delta U_{max}/U_{\infty})_{\textrm{Case 4}} \geq (\Delta U_{max}/U_{\infty})_{\textrm{Case 3}} \geq  (\Delta U_{max}/U_{\infty})_{\textrm{Case 2b}}$ for all streamwise positions, with more pronounced differences for low porosity discs, for which the levels of turbulence in the wake are closer to the ambient turbulence. As mentioned previously, because the turbulence generated by discs D4 and SD is significantly stronger than the ambient turbulence, the additional impact of FST is reduced, and observed primarily in the near wake. For these discs, the far wake turbulence and velocity deficit evolution are dominated by the turbulence generated by the discs themselves.

If we focus on the two FST “flavours” characterised by a low integral length scale (Group 1, which includes Case 1 and Case 2a), it can be observed in FIG.~\ref{Fig:MaxVeloDef} that the impact of increasing $TI_{\infty}$ on the evolution of $\Delta U_{max}/U_{\infty}$ varies depending on the disc porosity. For the solid disc, the trend is similar to the Group 2 cases, exhibiting a decrease in maximum velocity deficit for the wake experiencing higher turbulence intensity in the background (Case 2a). However, interestingly, for discs D2, D3, and D4, the maximum velocity deficit is higher for all streamwise positions in Case 2a with a higher $TI_{\infty}$ compared to Case 1. Hence, for these discs, increasing the intensity of a turbulence background with a very low integral length scale slows down the velocity recovery.

Further investigation would be necessary to draw conclusions regarding this non-linear effect of turbulence intensity at low integral length scale depending on the porosity of the disc. A study with a broader range of $TI_{\infty}$ with a very low $L_0$ would be valuable to further address this subject. However, at first glance, since different disc porosity creates a different near wake with different flow pattern (including the existence, position, size of the recirculation region in the wake, and the presence or absence of vortex shedding, see~ \citep[\emph{e.g.}][]{Castro1971,Xiao2013,Theunissen2019}), it is very likely that the influence of FST on wake recovery varies strongly with porosity, potentially exhibiting non-linear effects. For instance, vortex shedding downstream of porous discs only occurs above a specific porosity threshold \cite{Castro1971}. Consequently, depending on the presence or absence of these large-scale coherent structures in the wake, the effects of FST on the wake development may vary. This topic will be further explored in \S~\ref{subsection:vortex shedding}. However, overall, it is evident that the inflow integral length scale plays a significant role in the wake recovery behind porous discs, as is observed with wind turbines \cite{Gambuzza2022,Hodgson2023} or with solid bluff bodies \cite{Kankanwadi2023}. The effect of the intensity of the turbulent ambient on wake recovery is probably not independent of its integral length scale and, at least for porous discs, not uncorrelated from their thrust coefficient.

\subsection{ Entrainment characteristics }

In this subsection, we will analyse the entrainment characteristics of the background flow into the wakes from a mean flow perspective. The entrainment, in a time-averaged sense, can be described by considering the streamwise evolution of the wake width $\Delta \delta_{0.5}$ (\S~\ref{subsec:wake width}), and mass-flux $\dot{m}$ (\S~\ref{subsec:MassFlux}) \cite{Mistry2016,Breda2018,Xu2023}. 

\subsubsection{Wake width \label{subsec:wake width}}

\begin{figure}[tp!]
    \centering
    \includegraphics[width=\textwidth]{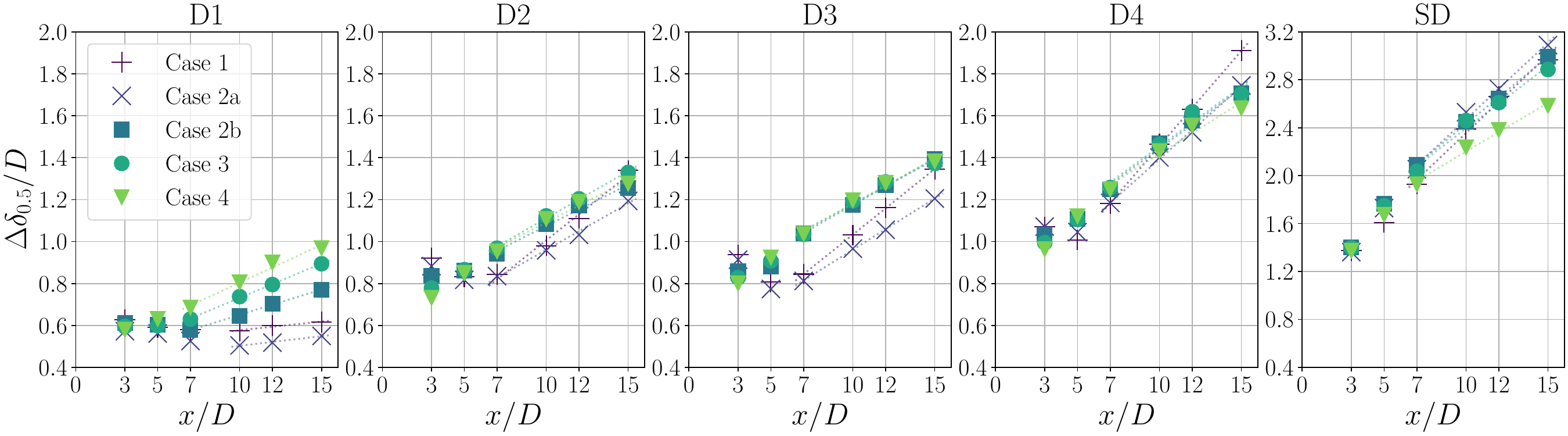}
    \caption{Evolution of the wake width $\Delta \delta_{0.5}/D$ with increasing downstream distance $x/D$. Note the change in the ordinate axis for SD.}
    \label{Fig:WakeWidth}
\end{figure}

FIG.~\ref{Fig:WakeWidth} displays the streamwise evolution of $\Delta \delta_{0.5} /D$ for all wakes, which is defined as the distance between the two dimensionless radial locations where the velocity deficit is half that of the maximum. Interestingly, for all four porous discs, both in the non-turbulent case 1 and in case 2a with a very low integral length scale (Group 1 cases), the wakes initially contract before widening. This initial shortening of the wakes has been reported not only with porous discs \citep[\emph{e.g.}][]{Vinnes2023} but also with both lab-scale \cite{Vinnes2022} and utility-scale wind turbines \cite{Dasari2019}. As the solidity of the disc increases, the transition point from contraction to expansion moves closer to the disc, and ultimately, is not observed for the solid disc within our current field of measurement. While the initial wake shrinking is still noticeable for D1 in the FST cases 2b and 3, it is no longer visible within our current field of view for the highest $TI_{\infty}$ and $L_0$ cases for D2, D3, and D4 (Group 2 cases). Hence, increasing the intensity or the integral length scale of the FST either suppresses the wake contraction region or shifts it closer to the discs, upstream of the measurements conducted in the present study. 

At every streamwise position $x/D$ and for all FST conditions, the wake width increases with the solidity/the thrust of the discs (note: the ordinate axis in FIG.~\ref{Fig:WakeWidth} varies between D4 and SD). For all porous discs, the wake width at the nearest measurement position $x/D = 3$ decreases as $TI_{\infty}$ or $L_0$ increases, particularly for D2, D3 and D4. Focusing on disc D1, with the highest porosity, and on the Group 2 cases (2b, 3 and 4), it can be observed that increasing the turbulence intensity of a turbulent background with a high integral length scale leads to a wider wake at all measurement stations. Conversely, in cases with a low integral length scale (Group 1 cases), an increase in $TI_{\infty}$ reduces the wake width. This trend is similarly observed with discs D2, D3 and D4; however for Group 2 cases, the difference in wake width is less pronounced compared to D1. This is likely due to the higher turbulence intensity generated by these discs, which potentially diminishes the additional impact of FST on wake expansion. These results once again underscore that the influence of $TI_{\infty}$ on wake development is contingent upon both the integral length scale of the background turbulence $L_0$ and the porosity $\beta$ of the discs. As porosity decreases, the difference in $\Delta \delta_{0.5}$ within the far wake between Group 2 cases and Group 1 cases gradually diminishes. Specifically, for discs D4 and SD, the wake at $x/D = 15$ is narrower for all Group 2 cases compared to Group 1 cases. Notably, the lowest $\Delta \delta_{0.5}/D$ is observed in Case 4 when the turbulence intensity of the background is at its highest level. Even though at the nearest measurement location, the wake is initially wider in the presence of FST, the wake width then grows slower when the background is turbulent.

\begin{figure}
        \centering
        \includegraphics[width=0.45\textwidth]{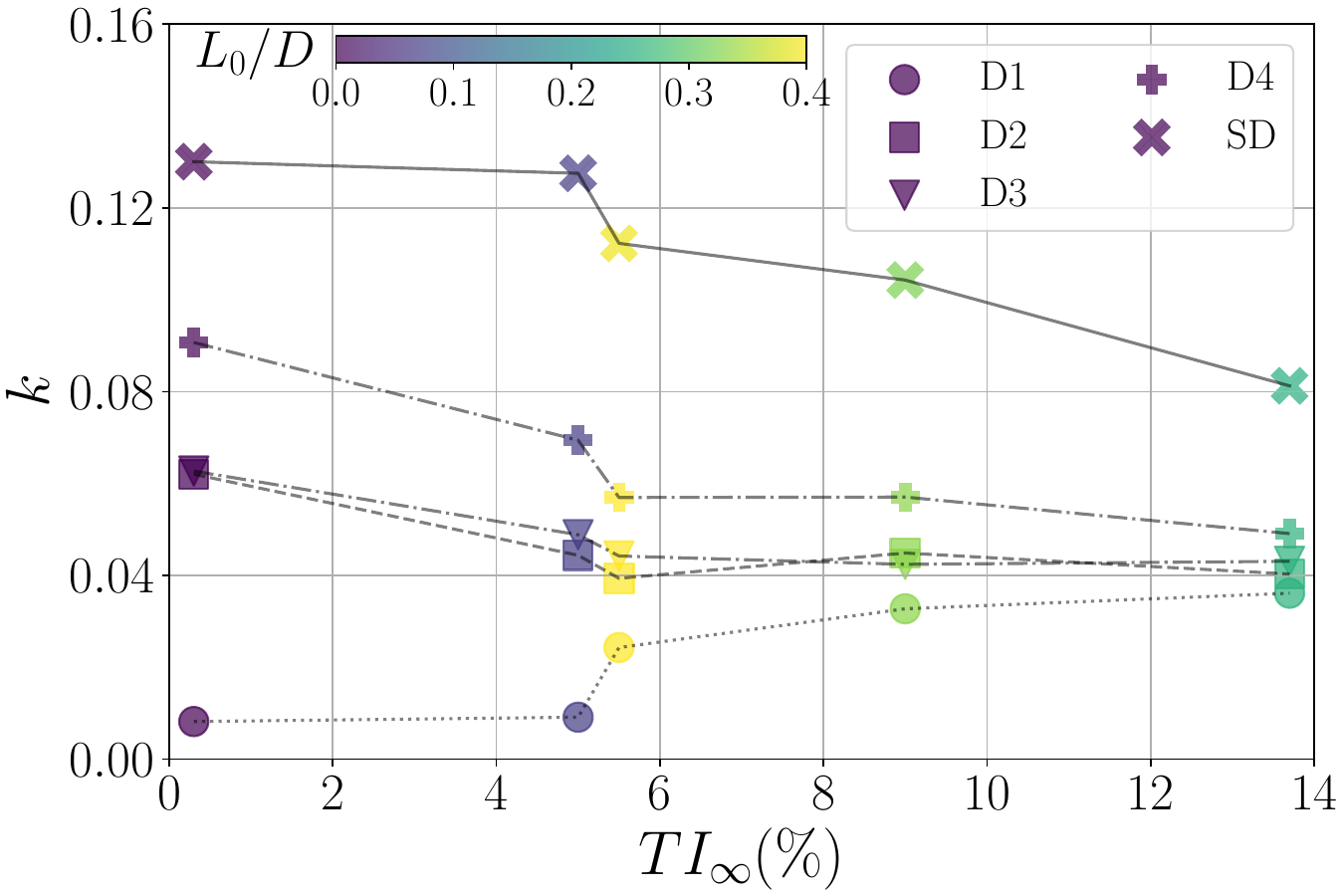}
        \caption{Wake growth rate $k$ for all $\{\textrm{disc}/\textrm{FST case}\}$ pairs plotted against the freestream turbulence intensity $TI_{\infty}$ (horizontal axis) and integral length scale $L_0/D$ (colour axis). The wake growth rate is defined as the slope of the linear regression of $\Delta \delta_{0.5} /D $ (see FIG.~\ref{Fig:WakeWidth}). Linear regressions for all $\{\textrm{disc}/\textrm{FST case}\}$ pairs were performed for $x/D \geq 7$, except for pairs $\{\textrm{D1}/\textrm{Cases 1 \& 2a}\}$, for which regressions were conducted for $x/D \geq 10$.}
        \label{Fig:WakeWidth_k}
\end{figure}

As is common practice for wind turbines wakes \citep[\emph{e.g.}][]{Bastankhah2014,PorteAgel2019,Ozturk2022,Gambuzza2022}, we performed linear fits of the wake width evolution and identified $k = \textrm{d}( \Delta  \delta_{0.5}) / \textrm{d} x$ as the wake growth rate. For all $\{\textrm{disc}/\textrm{FST case}\}$ pairs, linear regressions have been performed for $x/D \geq 7$, except for $\{\textrm{D1}/\textrm{Cases 1 \& 2a}\}$ pairs, where regressions were conducted for $x/D \geq 10$ to start fitting after the initial wake contraction. The growth rates of all wakes are presented in FIG.~\ref{Fig:WakeWidth_k} as a function of the ambient turbulence intensity $TI_{\infty}$ (abscissa) and integral length scale $L_0$ (colour bar).

The range of wake growth rates observed with the discs is representative of wind turbine wakes, where values typically range between $k=0.01$ and $k=0.15$, depending on the thrust coefficient and the FST conditions. \cite{Ozturk2022,Gambuzza2022}. The effect of the disc porosity is evident, showing an increase in the wake growth rate as the porosity decreases, \emph{i.e.} as the thrust coefficient increases. The trend regarding the effect of $C_T$ on wind turbine's wake growth rate in the literature is erratic, with some papers reporting a decrease with $C_T$ \citep[\emph{e.g.}][]{Ozturk2022}, while others indicate an increase \citep[\emph{e.g.}][]{Ishihara2018,Gambuzza2022}. Moreover, in a wide majority of wind turbine wake models, the impact of the thrust coefficient on wake width is typically solely accounted in the initial wake width, and not in the wake growth rate, under the assumption that higher thrust results in a wider initial wake. We observed a similar trend with porous discs, but additionally, we demonstrate that the thrust coefficient significantly affects the wake growth rate $k$, at least for porous discs.

Interestingly, the influence of FST on the wake growth rate varies depending on the disc porosity and thrust coefficient. For D1, the presence of FST increases the wake growth rate $k$, with noticeable effects from both turbulence intensity $TI_{\infty}$ and integral length scale $L_0$. However, for all three other porous discs and the solid disc, the effect of FST on the wake growth rate is opposite, with a clear reduction in the spreading rate in the presence of FST. The highest wake growth rates are observed in Case 1, without FST, while the lowest rates are observed in Case 4, where the turbulence intensity of the ambient flow is at its highest level. Faster expansion of wakes in a non-turbulent ambient compared to a turbulent ambient has also been reported in studies involving bluff bodies~\cite{Kankanwadi2020} and non-uniform porous discs \cite{Vinnes2023}. Furthermore, regarding the impact of the turbulent integral length scale $L_0$, all Group 2 cases exhibit lower $k$ values compared to Group 1 cases. Additionally, there is a clear reduction in $k$ when comparing Cases 2a and 2b, which are characterised by the same $TI_\infty$, but with the latter case having an integral length scale five times larger. Hence, it can be concluded that both $TI_{\infty}$, $L_0$, and $\beta$ play crucial role in determining the wake growth rate. Below a certain threshold of porosity, the presence of FST leads to a decrease in the wake growth rate, with noticeable impacts from both turbulence intensity and integral length scale. However, beyond this threshold, the effect of FST reverses. 

In most wind turbine wake models, the wake growth rate $k$ is commonly assumed to follow a linear or power function of the turbulence intensity $TI_{\infty}$, with no dependency on either the thrust coefficient produced by the turbine $C_T$ or the length scale of the ambient turbulence $L_0$. However, we found that both $L_0$ and $C_T$ play crucial roles in determining this parameter. Moreover, similar to discs, the effects of $TI_{\infty}$ on the wake growth rate may vary for a wind turbine depending on $L_0$, $C_T$ and the distance in the wake from the turbine.

\subsubsection{Mass flux and entrainment rate \label{subsec:MassFlux} }

\begin{figure}[b]
    \centering
    \includegraphics[width=\textwidth]{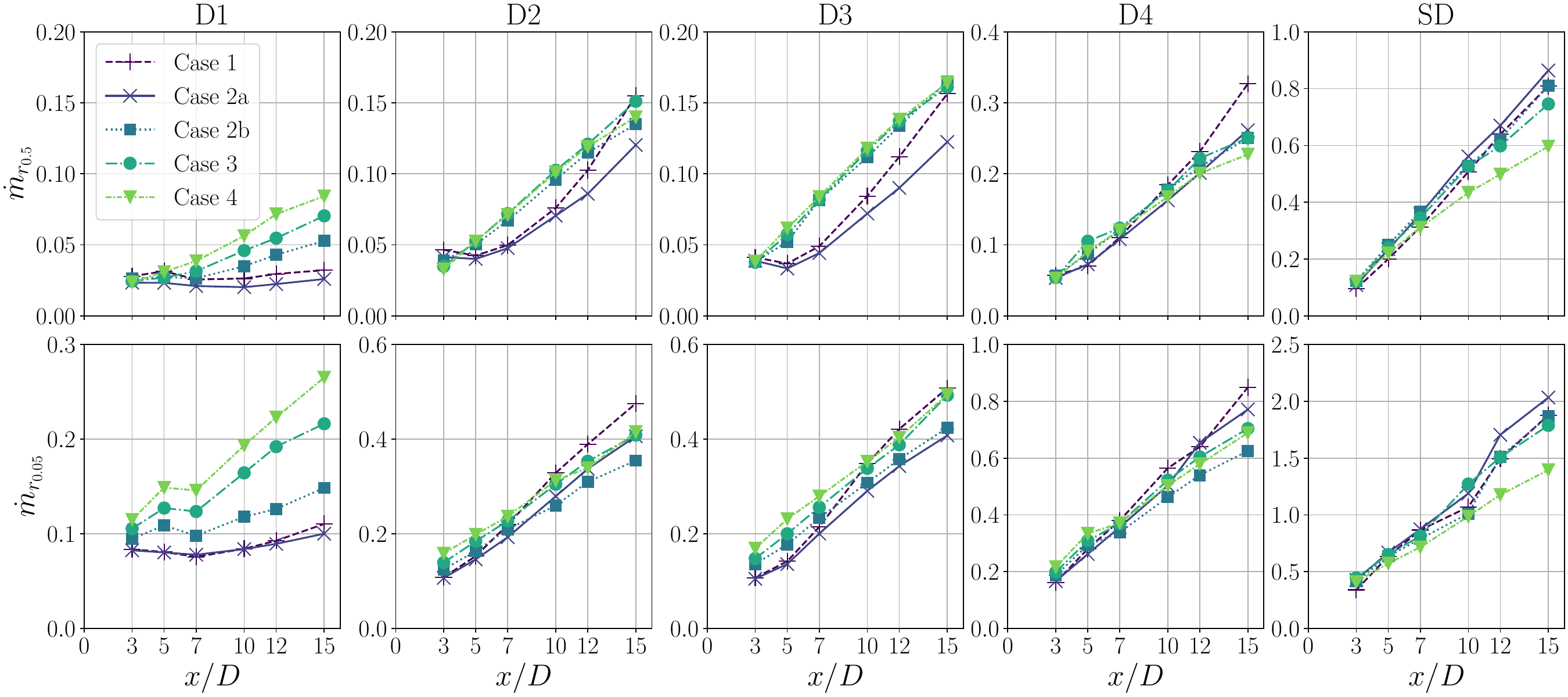}
    \caption{Evolution of the global integral mass-flux $\dot m$ with increasing downstream distance $x/D$. Top row: Mass-flux $\dot m_{r_{0.5}}$ computed up to the wake half-width $\delta_{0.5}$, defined as the radial position $r_{0.5}$ where $\Delta U = 0.5 \Delta U_{max}$. Bottom row: Mass-flux $\dot m_{r_{0.05}}$ computed up to the radial position $r_{0.05}$ where $\Delta U = 0.05 \Delta U_{max}$. Note the change in the ordinate axis across the various discs.}
    \label{Fig:IntegralMassFlux}
\end{figure}

\begin{figure}[btp]
    \centering
     \begin{minipage}[t]{0.45\textwidth}
        \includegraphics[width=\textwidth]{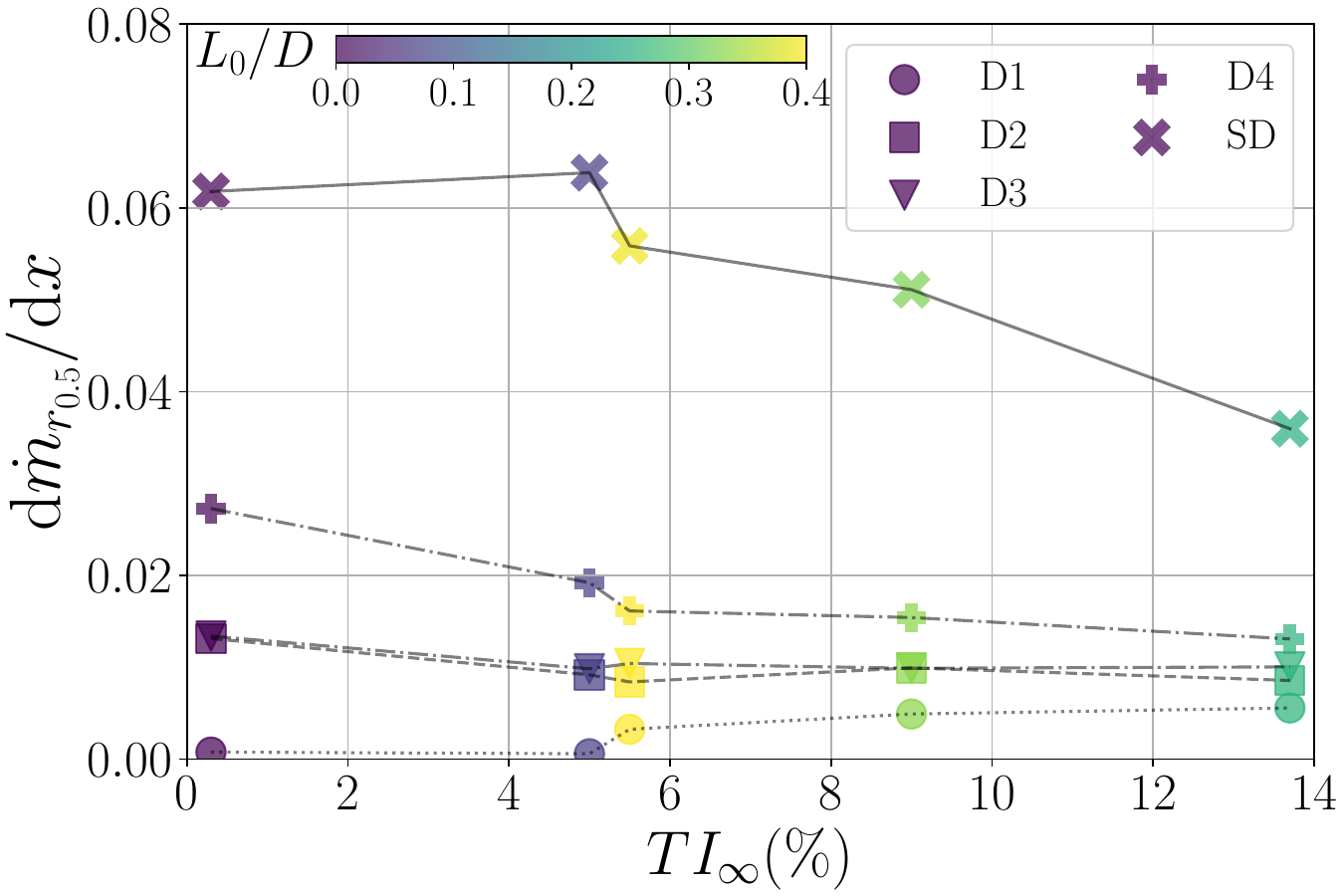}
        \subcaption{}
        \label{Fig:MassFluxRate05}
    \end{minipage}
    \hfill
    \begin{minipage}[t]{0.45\textwidth}
        \centering
        \includegraphics[width=\textwidth]{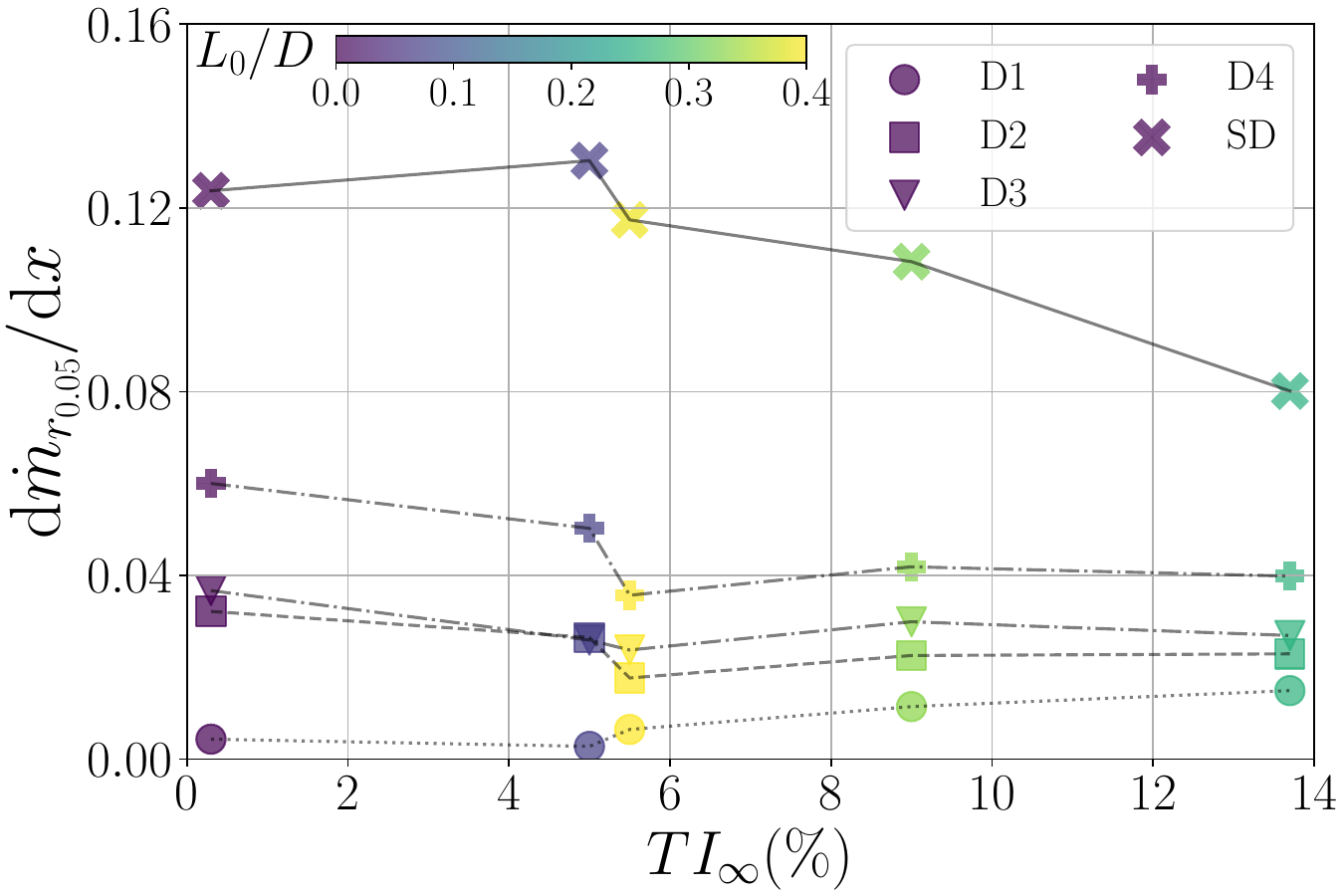}
        \subcaption{}
        \label{Fig:MassFluxRate005}
    \end{minipage}
    \caption{Integral mass-flux rate $\textrm{d}\dot m /\textrm{d}x $ for all $\{\textrm{disc}/\textrm{FST case}\}$ pairs as a function of the freestream turbulence intensity $TI_{\infty}$ (horizontal axis) and integral length scale $L_0/D$ (colour axis). The integral mass-fluxes $\dot m_{r_{0.5}}$  (\subref{Fig:MassFluxRate05}) and $\dot m_{r_{0.05}}$ (\subref{Fig:MassFluxRate005}) are respectively computed up to $r_{0} = r_{0.5} = \delta_{0.5}$, and $r_{0} = r_{0.05}$.}
    \label{Fig:Entrainemnt rate}
\end{figure}

In order to gain further insight on the entrainment characteristics, we evaluated the streamwise evolution of the global integral mass-flux $\dot m$ computed as 
\begin{align}
    \dot m = 2 \pi \rho \int_0^{r_0} U r dr
\end{align}
The integral mass-flux has been computed up to two different integral boundaries $r_0$. On one hand, we computed the integral mass-flux in the “central wake region", denoted as $\dot{m}_{r_{0.5}}$, where $r_0$ corresponds to the wake half-width ($r_0 = \delta_{0.5}$). On the other hand, we calculated a second integral mass-flux encompassing a broader region of the wake, denoted as $\dot{m}_{r_{0.05}}$, with $r_0$ being the radial position where the mean velocity deficit $\Delta U$ is $5\%$ of the maximum velocity deficit $\Delta U_{max}$. This second integral boundary $r_0$ has been selected to be as large as possible (thus $\Delta U$ is small) considering the experimental noise inherent in hot-wire measurements. Similarly to Mistry~\emph{et al.}~\cite{Mistry2016}, the overall mass entrainment rates were then determined from the streamwise gradient of the mass fluxes, $\textrm{d} \dot m / \textrm{d}x $, computed for all cases from $x/D \geq 7$, except for the configurations $\{\textrm{D1}/\textrm{Case 1}\}$ and $\{\textrm{D1}/\textrm{Case 2a}\}$, for which the the mass entrainment rates were computed for $x/D \geq 10$, to exclude the the initial wake contraction.

FIG.~\ref{Fig:IntegralMassFlux} and FIG.~\ref{Fig:Entrainemnt rate} respectively show the streamwise evolution of the integral mass fluxes ($\dot{m}_{r_{0.5}}(x)$ and $\dot{m}_{r_{0.05}}(x)$), and the entrainment rates ($ \textrm{d}\dot{m}_{r_{0.5}}/\textrm{d}x$ and $\textrm{d}\dot{m}_{r_{0.05}}/\textrm{d}x$) for all $\{\textrm{disc}/\textrm{FST case}\}$ combinations. Firstly, selecting $r_{0.5} = \delta_{0.5} = r(\Delta U = 0.5 \Delta U_{\text{max}})$ or $r_{0.05} = r(\Delta U = 0.05 \Delta U_{\text{max}})$ as a boundary for the integral mass-flux does not alter the trend of both the evolution of the mass-flux and the entrainment rate with respect to porosity or the FST conditions, but only affects their magnitudes. As expected, the evolution of the mass-flux in the “central wake region", $\dot{m}_{r_{0.5}}$, closely mirrors the evolution of the wake width (FIG.~\ref{Fig:WakeWidth}). Specifically, we can observe a change in the slope of the inner wake mass-flux for D1, D2 and D3 in cases 1 and 2a around $x/D \approx 7$, similar to the change in the slope of the wake width evolution, \emph{i.e.} the transition from wake contraction to expansion. Except for D1, we do not observe this initial reduction in the mass-flux when integrating up to $r_{0.05}$.

Decreasing the porosity leads to an increase in the mass-flux and the entrainment rate, which is consistent with the increase in the wake width and wake growth rate discussed previously. At the furthest distance from the discs ($x/D = 15$), the integral mass-flux in Group 1 cases (low-turbulence inflow) is generally higher than in Group 2 cases (high-turbulence cases) for all discs, except D1. For $x/D \geq 7$, the entrainment rates for D2, D3, D4 and SD decrease in the presence of FST, with clear effects of both the turbulence intensity and integral length scale of the background flow. This finding supports the results reported by Kankanwadi \& Buxton~\cite{Kankanwadi2020}, who did observe a reduction of the mass entrainment flux in the far wake of a circular cylinder in the presence of FST. However, closer to the discs, the mass fluxes are higher for Group 2 turbulence “flavours”, at least for discs D1, D2, D3 and D4. This change in how FST affects the mass-flux might be related to a transition of the wake growth entrainment mechanisms from being driven by both nibbling and engulfment to being solely driven by nibbling. Indeed, in a wake where large eddies are energetic, such as the near wake of a solid bluff body, engulfment is significant in the entrainment process and wake growth. In such wakes, Kankanwadi \& Buxton~ \cite{Kankanwadi2023} reported that FST enhances the wake growth. However, when nibbling dominates the entrainment process, such as in the far wake of a solid bluff body, freestream turbulence intensity in the background suppresses nibbling, and thus decreases the mean entrainment mass-flux \cite{Kankanwadi2020}. To gain deeper insights into the entrainment mechanisms for the various wakes generated in the current study, a more thorough analysis of the structures within the wakes is carried out in the following section. As porosity increases, the FST effect observed by Kankanwadi \& Buxton with a solid body diminishes; \emph{i.e.} the freestream turbulence intensity has a greater effect on decreasing the entrainment rate in the far wake, and actually the wake growth rate, for SD and D4 than it does for D3 and D2. Furthermore, D1 actually shows an opposite trend of increasing entrainment rate and wake growth rate with freestream turbulence intensity. Hence, the entrainment into the wake of a highly porous object exhibits the opposite trend compared to a solid object. The very low turbulence intensity added by D1 into the flow, comparable to the freestream turbulence intensity, as well as the very small velocity deficit relative to the r.m.s of the freestream velocity in turbulent cases might play a role in this observed change in trend.

\subsection{Fluctuating velocity spectra \label{subsection:vortex shedding}}

\begin{figure}[tp]
    \centering
     \begin{minipage}[c]{0.47\textwidth}
        \includegraphics[width=\textwidth]{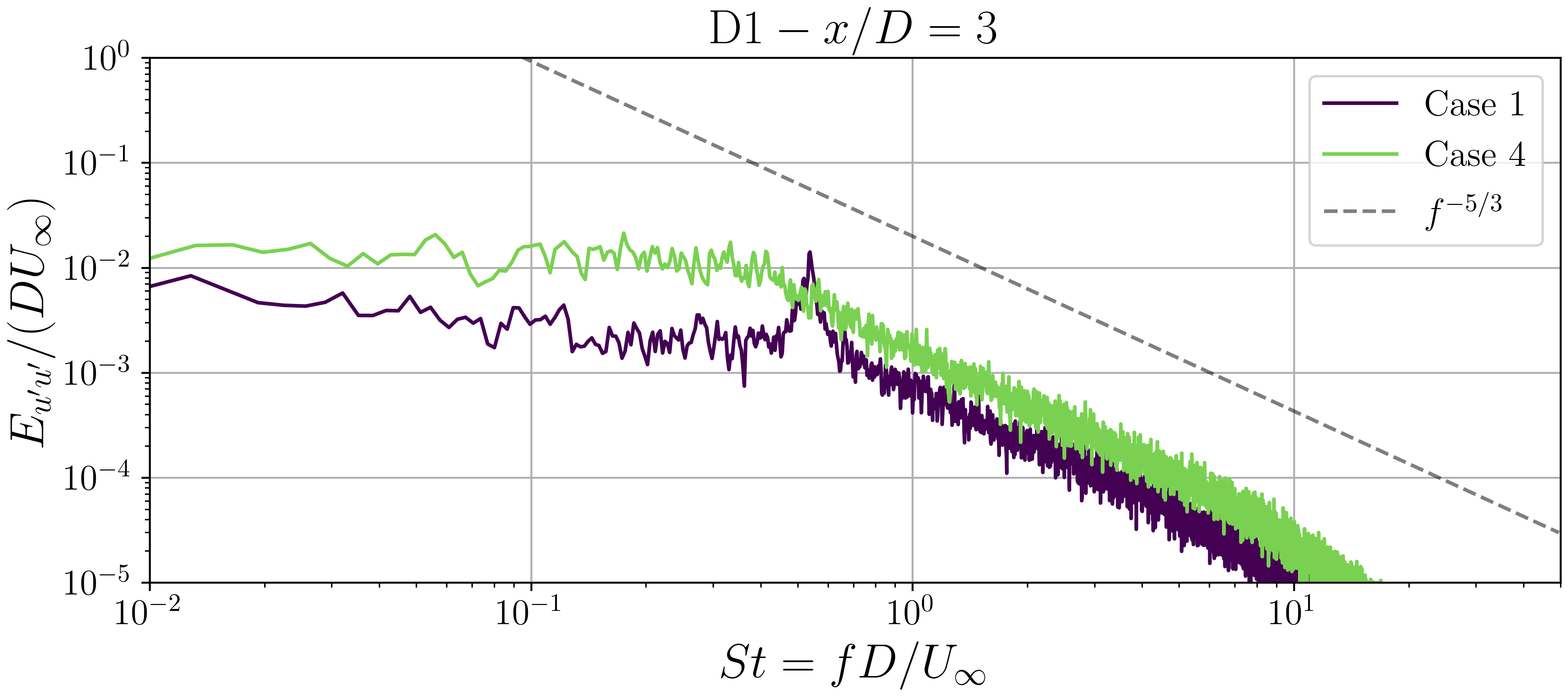} \\
        \includegraphics[width=\textwidth]{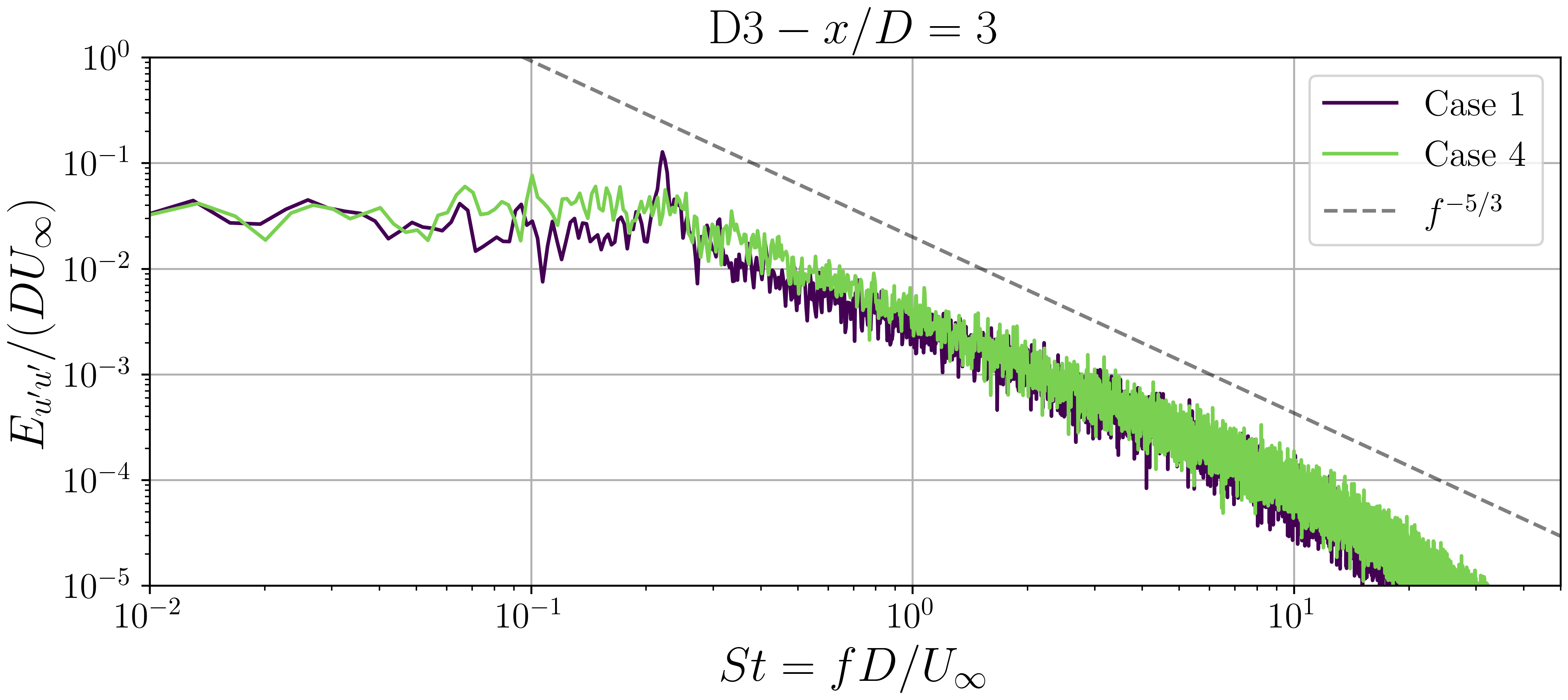}\\
        \includegraphics[width=\textwidth]{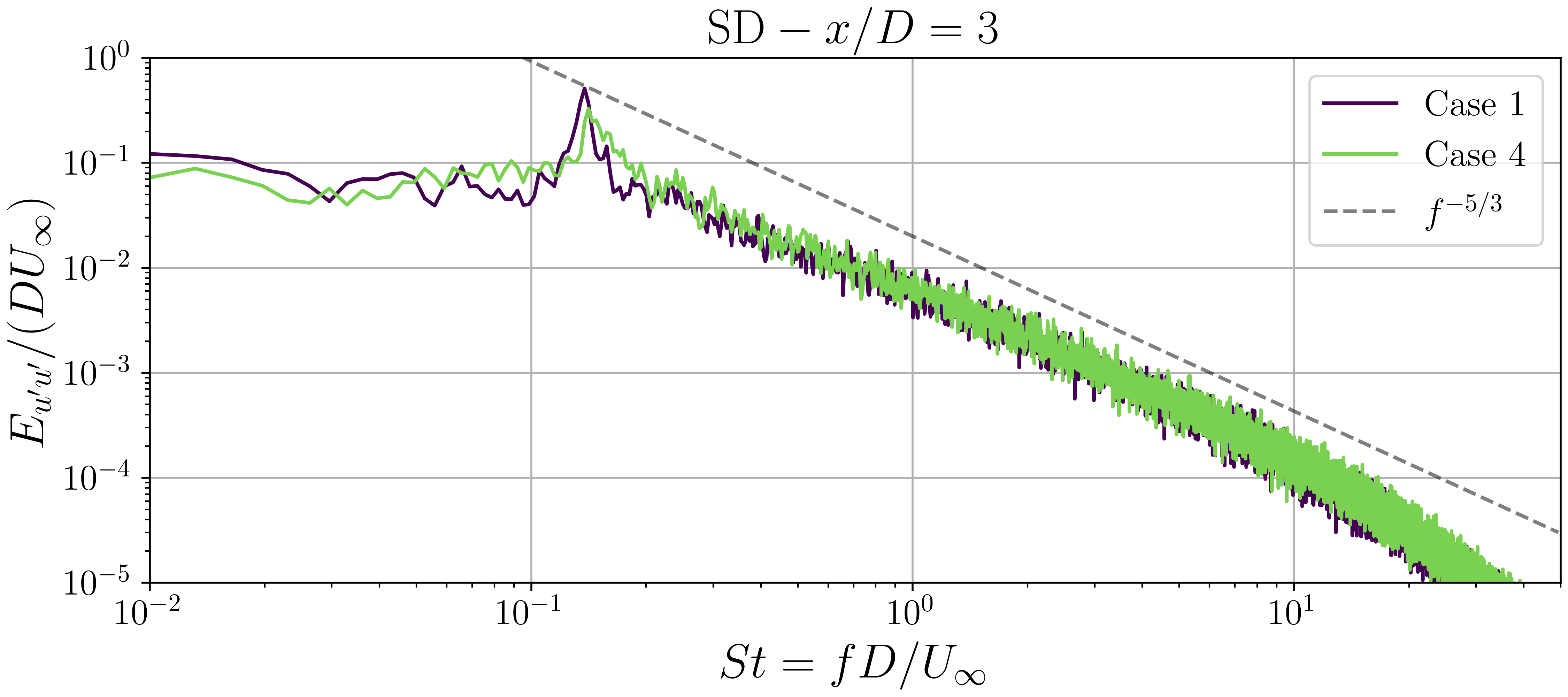}
        \caption{Power spectral density of the velocity fluctuations $E_{u'u'}/(DU_{\infty})$ plotted against the Strouhal number $St$ for D1 (top), D3 (middle) and SD (bottom) in Cases 1 and 4. All spectra are taken at $x/D = 3$, and at the spanwise position $y/D$ where the peak frequency attains the highest magnitude.}
        \label{Fig:Spectra}
    \end{minipage}
    \hfill
    \begin{minipage}[c]{0.47\textwidth}
        \centering
        \begin{minipage}[c]{\textwidth}
        \centering
        \includegraphics[width=\textwidth]{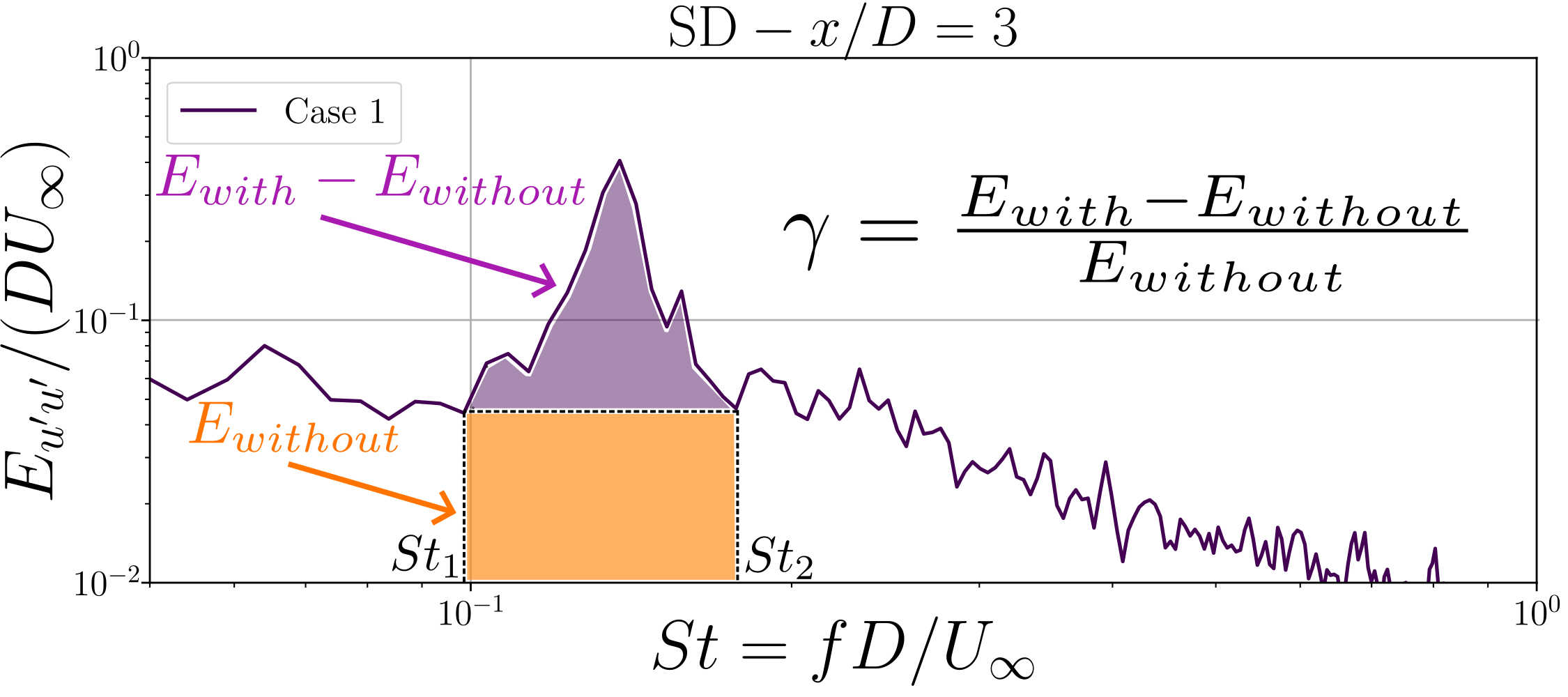}
        \caption{Schematic illustrating the calculation of the parameter $\gamma$. The velocity spectrum at $x/D = 3$ and $y/D =0.5$ in the wake of the configuration  $\{\textrm{SD}/\textrm{Case 1}\}$ is used for this illustration.}
        \label{Fig:RindCastro}  
        \end{minipage} \\ \vspace{0.8cm}
        \begin{minipage}[c]{\textwidth}
        \centering
        \includegraphics[width=\textwidth]{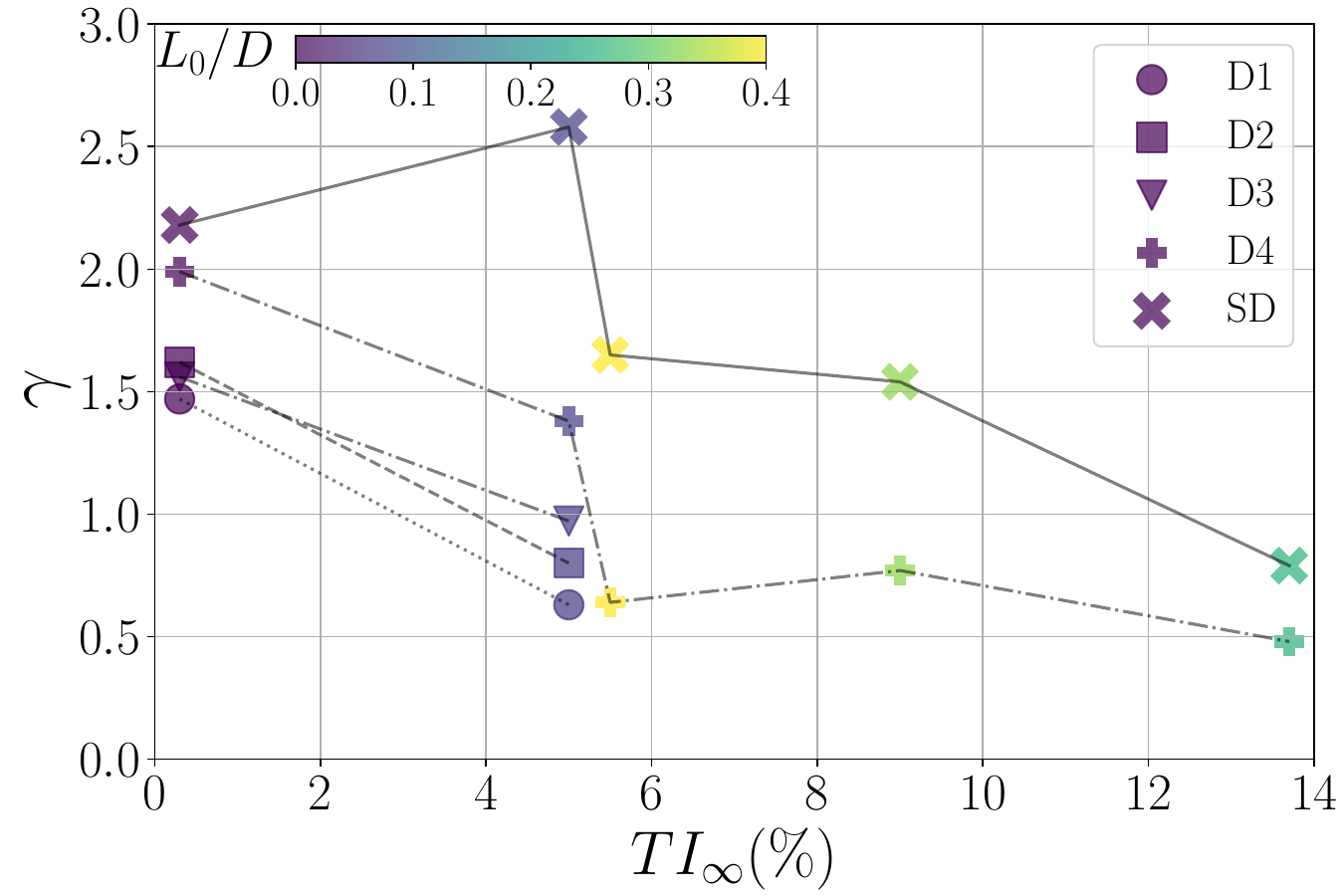}
        \caption{Parameter $\gamma$ for all $\{\textrm{disc}/\textrm{FST case}\}$ pairs plotted against the freestream turbulence intensity $TI_{\infty}$ (horizontal axis) and integral length scale $L_0/D$ (colour axis).}
        \label{Fig:Gamma}    
        \end{minipage}
    \end{minipage}  
\end{figure}

To obtain further information about the structures within the different wakes, we calculated the power spectral density of the velocity fluctuations, denoted here as $E_{u'u'}$. Some normalised velocity spectra $E_{u'u'}/(DU_{\infty})$ downstream of D1, D3 and SD are displayed in FIG.~\ref{Fig:Spectra} for two different FST “flavours” (Case 1 and Case 4). The abscissa corresponds to the Strouhal number based on the disc diameter and freestream wind velocity, $St = fD/U_{\infty}$. 

A distinct low-frequency peak is observed in the velocity spectra at $x/D=3$ for certain $\{\textrm{disc}/\textrm{FST case}\}$ combinations. The corresponding Strouhal numbers are displayed in TABLE~\ref{tab:strouhal}. In all FST cases, distinct peaks are observed in the SD spectra. In the absence of FST, the Strouhal number $St=0.138$ associated to this peak closely matches the expected $St$ related to the standard vortex shedding downstream of a solid disc (\emph{e.g.} $St =0.137$ in \cite{Vinnes2022}, $St =0.14$ in \cite{Bearman1983}, $St =0.13$ in \cite{Cannon1993}). Consistent with the findings reported by Rind \& Castro \cite{Rind2012}, we observe a slight increase in the vortex-shedding frequency as the freestream turbulence intensity and integral length scale increase. Similarly, in all FST cases, distinct peaks around $St = 0.14$ are observed in the D4 spectra, indicating the presence of standard vortex shedding downstream of this disc. For both D4 and SD, these peaks persist in the velocity spectra up to $x/D = 15$ in all FST cases, albeit with diminishing magnitude, reflecting the gradual decay of the large-scale coherent motions in the wake.

For the three other porous discs (D1, D2 and D3), low-frequency peaks are only observed in the two low-turbulence cases 1 and 2a (Group 1 cases). The Strouhal numbers of the peaks increase as the porosity increases ($St = 0.221$ for D3, $St = 0.272$ for D2, and $St = 0.538$ for D1 in Case 1) and, similarly to the solid disc, increase as the turbulence intensity of the background increases. Interestingly, even with the most porous disc D1 ($\beta = 0.6$), we observe a clear peak in the velocity spectra in Case 1 (FIG.~\ref{Fig:Spectra}). This contrasts with the results reported by Cannon \emph{et al.}~\cite{Cannon1993}, who did not observe vortex shedding downstream of wire mesh porous discs with a porosity greater than 0.4. However, the discs in the present study have a non uniform porosity and a solid centre to simulate the presence of the nacelle. Therefore, it is plausible that the peak observed in the spectra of D1 is associated with the shedding of the high-solidity central region of the disc. Indeed, the transverse location $y/D$ where the energy content of the low-frequency band in the velocity spectra is maximum $(E_{u'u'}/(DU_{\infty}))_{max}$ is typically close to the disc's edge, within the shear layer ($y/D \approx 0.5D$), for all discs except D1. For D1, the energy content of the low-frequency peak is maximum at $y/D = 0.14$. Then, if we compute an alternative Strouhal number for D1, $St' = fD'/U_{\infty}$, based on a representative diameter of this low-porosity region rather than on the disc diameter, \emph{e.g.} $D' = 0.14D/0.5D$, we find a Strouhal number close to $St' = 0.15$, representative of the vortex shedding of a solid disc. Hence, it is probable that the low-frequency peaks in the D1 spectra are associated with the vortex shedding of the low-porosity central disc region, which simulates the nacelle. However, further investigation would be required to ascertain whether the low-frequency peaks observed in the velocity spectra of the three most porous discs are associated with regular vortex shedding, similar to what is observed with SD and D4, or if these periodic effects are linked to some sort of far wake instability \cite{Castro1971}.

\begin{table}[t]
\caption{\label{tab:strouhal}%
Strouhal numbers $St = fD/U_{\infty}$ for all $\{\textrm{disc}/\textrm{FST case}\}$ configurations.}
\begin{ruledtabular}
\begin{tabular}{c c c c c c }
& Case 1 & Case 2a& Case 2b & Case 3& Case 4 \\
D1 & 0.538  &  0.549 & -&- &- \\
D2 & 0.272   & 0.286  & - & -&- \\
D3 & 0.221 & 0.250 & - & - & - \\
D4 & 0.142 & 0.138 & 0.141 & 0.134 & 0.136 \\
SD & 0.138 & 0.139 & 0.148 & 0.145& 0.144 
\end{tabular}
\end{ruledtabular}
\end{table}

Interestingly, for the three discs with the highest porosity (D1, D2, D3), we do not detect clear peaks at low Strouhal numbers for the three highly turbulent cases (Group 2 cases), indicating a significant effect of the freestream turbulence on the energy content of these low frequencies. The effect of disc porosity and FST on the energy content of these low-frequency peaks is evaluated following the same approach as Rind \& Castro~\cite{Rind2012}. In this aforementioned study, Rind \& Castro introduced a parameter $\gamma$, defined as
\begin{align}
    \gamma  &=\frac{E_{with} - E_{without} } {E_{without}}
\end{align}
where $E = \int_{St_1}^{St_2} E \textrm{d}St$ is the energy content of a range of frequencies bounding the vortex shedding peak, with the subscripts \textit{with} and \textit{without} respectively indicating whether the peak is included ($E_{with}$) or excluded ($E_{without}$) from the integral. A sketch, mirroring that of Rind \& Castro~\cite{Rind2012}, which illustrates this definition, is presented in FIG.~\ref{Fig:RindCastro}. $\gamma$ provides an estimate of the energy contribution of the low-frequency broad peak to the full spectrum, thereby allowing for an estimation of the relative strength of the vortex-shedding process (at least for SD and D4, for which the peaks correspond to regular vortex shedding). FIG.~\ref{Fig:Gamma} shows $\gamma$ for $\{\textrm{disc}/\textrm{FST case}\}$ combinations where a clear low-frequency peak is observed at the nearest position to the discs, $x/D=3$. $\gamma$ has been calculated at the horizontal position $y/D$ where the energy content of the peak is maximal. In the near wake, it is clear that both an increase in porosity or the presence of FST decreases the relative energy that the shedding process contributes to the spectrum, with also clear effects of both turbulence intensity $TI_{\infty}$ and integral length scale $L_0$. We can hypothesize that FST either weakens the initial energy content of vortex shedding or accelerates the decay of the large-scale coherent structures in the wake, or possibly both. Combined with the fact that the energy content of the large-scale coherent structures decays with downstream distance, it also explains why this low-frequency component was not detected in the spectra of Group 2 cases after a certain downstream distance, in particular for the most porous discs. For instance, for D1, the peak vanishes in the wake region at $x/D = 7$ in case 1 and $x/D = 5$ in case 2a.

From an entrainment perspective, large-scale coherent structures play a crucial role in the near-wake mixing process and the wake growth downstream of a solid bluff body. Therefore, we will now analyse the entrainment characteristics in light of the velocity spectra. Firstly, it is noteworthy that the initial narrowing of the wake downstream of the porous discs in the two low $\{TI_{\infty},L_0 \}$ cases (Group 1 cases, see FIG.~\ref{Fig:WakeWidth}) coincides with the presence of the low-frequency broad peak in the velocity spectra (for D1, D2, and D3), or coincides with the cases for which $\gamma$ is maximum (for D4). In this near wake region ($x/D \leq 7$), wakes are wider and grow faster for the Group 2 cases (high $\{TI_{\infty},L_0 \}$ cases) compared to the Group 1 cases (low $\{TI_{\infty},L_0 \}$ cases) (FIG.~\ref{Fig:WakeWidth}). We can therefore postulate that, similarly to wind turbine tip vortices, the large-scale structures shed by the porous discs act as a shield against mixing, preventing the background flow from being “entrained" into the wake and inhibiting its initial growth. As the addition of turbulence suppresses, or at least weakens, the vortex-shedding process (see FIG.~\ref{Fig:Gamma}), it is consistent that we observe wider wakes that expand faster in the presence of FST in the near wake region.

Due to the absence of clear low-frequency peaks in the velocity spectra, and hence absence of large-scale coherent motions, for D1, D2, and D3 in the Group 2 cases, nibbling might be the dominant entrainment mechanism across the entire measurement range for these three FST cases. However a transition in the entrainment process from being significantly influenced by engulfment to being solely driven by nibbling may occur in the Group 1 cases for these three discs, as well as for all 5 FST cases for D4 and SD. For the porous discs, we can hypothesise that after the initial shrinking of the wake occurring in Group 1 cases ($x/D \approx 10$ for D1, $x/D \approx 7$ for D2 and D3, and $x/D \approx 5$ for D4), the influence of large-scale coherent structures on wake expansion, and consequently entrainment by engulfment, becomes less significant. Indeed, beyond these points, the wake width grows faster, and the entrainment rate is higher in the non-turbulent case for D2, D3, and D4  (FIG.~\ref{Fig:WakeWidth_k} \& FIG.~\ref{Fig:Entrainemnt rate}). These characteristics are indicative of a wake that grows through nibbling-driven entrainment~\cite{Kankanwadi2020,Kankanwadi2023,Chen2023}. 

Similarly, for the solid disc, the entrainment rate and wake growth rate computed for $x/D \geq 7$ decrease in the presence of FST. Thus, we can hypothesise that, at least beyond this point, the energy content of the large-scale coherent structures has sufficiently decayed, thereby limiting their influence on the wake growth. Moreover if we focus on the $\{\textrm{SD}/\textrm{Case 4}\}$ combination, there is an apparent turning point of the slope of $\Delta \delta_{0.5}(x)$ and $\dot m(x)$ located at $x/D \approx 7$, after which the wake grows noticeably more slowly. This phenomenon is also observable in other FST cases, although the change in slope is less pronounced. Chen \& Buxton~\cite{Chen2023} observed this turning point in the wake of a solid circular cylinder, and reported that it corresponds to the location where the growth of the wake transitions from being large-scale engulfment-driven entrainment to small-scale nibbling-driven entrainment. Hence, we can postulate that the turning point located in the wake at $x/D \approx 7$ for the  $\{\textrm{SD}/\textrm{Case 4}\}$ pair marks the transition point at which large-scale engulfment likely subsides to negligibility. After reaching this point, we found a similar trend of reducing entrainment rate in the presence of FST.

For D1, D2, and D3, this turning point does not exist in Group 2 cases because of the absence of clear vortex shedding, while the turning point in Group 1 cases manifests as a transition from wake contraction to expansion. For the porous disc with the lowest porosity, D4, in low-turbulence cases 1 and 2a  (Group 1 FST cases), the behaviour is similar to that of the other porous discs, with an initial contraction of the wake followed by a fast expansion. In high-turbulence conditions (Group 2 FST cases), velocity spectra show peaks associated with vortex shedding, particularly in the near wake region. However, these peaks exhibit lower relative strength compared to those observed for the solid disc. Consequently, a turning point may also exist in the wake of D4, albeit with a less pronounced change in slope compared to the solid disc (as not clearly depicted in FIG.~\ref{Fig:WakeWidth}), or it might occur closer to the disc. Another possibility is that the large eddies shed by D4 are not sufficiently energetic to cause a significant alteration in the entrainment mechanism due to their energy decay, and thus may not significantly influence the slope of the wake width evolution.

In summary, we found that the wake width, wake growth rate, and entrainment rates are strongly correlated to the presence, absence, and strength of large-scale coherent structures within the wakes. FST and disc porosity have a strong effect on the initial energy content of these structures and their decay, which consequently affects the entrainment rate and wake development.

\section{Wake modelling - Application of the Townsend-George Theory \label{section: wake modelling}}

A thorough understanding of the evolution of the centreline mean velocity deficit $(\Delta U/U_{\infty})_{max}$ in the wake of a wind turbine is essential for optimising wind farm layout and control strategies. Hence, in recent years, significant effort has been dedicated to predicting the evolution of this quantity for a single turbine, resulting in the proposition of numerous wind turbine wake models (\citep[see \emph{e.g.}][]{Gocmen2016,PorteAgel2019} for reviews on wind turbine wake models).

Among the plethora of “engineering" models available, Gaussian-type models, such as the Bastankhah-Porté-Agel model \cite{Bastankhah2014}, and its derived versions \cite{Vahidi2022,Ishihara2018}, are among the most widely used wake models, supported by numerous numerical and experimental data \cite{Chamorro2009,Wu2012,Ishihara2018}. These analytical turbine wake models are derived from the fundamental governing equations, incorporating different assumptions such as the self-similarity and axisymmetry of the velocity profiles in the far wake, a linear growth rate of the wake, as well as the conservation of mass and momentum. Gaussian-type models predict a power law dependence of the centreline mean velocity deficit : $\Delta U_{max} \sim x^{-1}$.

On the other hand, the axisymmetric turbulent far wake of a solid bluff body in uniform inflow is a canonical turbulent free shear flow for which analytical models relying on fundamental and robust assumptions exist \cite{Townsend1976,George1989}. These analytical models rely on the self-similarity of the one-point turbulence statistics and the nature of the energy cascade, \emph{i.e.} how the energy is injected at large-scales and is dissipated at smaller scales. In the classical formulation introduced by Townsend \cite{Townsend1976} and George \cite{George1989}, the energy cascade is in equilibrium, with the decay of the power spectral density of the fluctuating velocity $E_{u'u'}$ following the Richardson–Kolmogorov $-5/3$ power law in the inertial sub-range. In this case, the dissipation rate ($\epsilon$) of the turbulent kinetic energy along the centreline ($K$) scales as $\epsilon = C_{\epsilon} K^{3/2}/L$ with $C_{\epsilon}$ being constant ($L$ is an integral length scale of the turbulence). The equilibrium prediction for the streamwise evolution of the the centreline mean velocity deficit for axisymmetric turbulent wakes developing in a uniform flow is then  $\Delta U_{max} \sim (x - x_0)^{-2/3}$. $x_0$ is a virtual origin that arises from the governing equations, and can be seen as an indication of the near wake length, even though the underpinning physics that control this value are not well understood \cite{Neunaber2024}. Recently a very different dissipation scaling, referred to as the non-equilibrium dissipation law, has been found, including in the decaying turbulence generated by fractal grids, with  $C_{\epsilon} = Re_G^m/Re_L^n\neq \textrm{const.}$ \cite{Dairay2015,Vassilicos2015,Obligado2016}. $Re_G$ is a global Reynolds number, and $Re_L$ a local turbulent Reynolds number. Different dissipation scalings of $\epsilon$ within a turbulent wind turbine wake would imply different scaling laws for $\Delta U_{max}$, and therefore, significant implications on the modelling of wind turbine wakes. For a turbulent axisymmetric wake produced by a solid object, it has been experimentally found that $m \approx n \approx 1$, leading to the the scaling law $\Delta U_{max} \sim (x - x_0)^{-1}$. Stein and Kaltenbach~\cite{Stein2019} found another couple of exponents $m \approx n \approx 2$ for a wake generated by a wind turbine and developing in a turbulent boundary layer, leading to another scaling law:  $\Delta U_{max} \sim (x - x_0)^{-2}$. TABLE~\ref{tab:scaling laws} displays the different theoretical axial scaling laws for $\Delta U_{max}$. 

Neunaber~\emph{et al.} examined the application of the Townsend-George theory to field measurement \cite{Neunaber2021_Townsend} and wind tunnel measurement \cite{Neunaber2022} of wind turbine wakes. They reported that wind turbine wakes partially fulfil the requirements under which the Townsend-George theory is valid, and found that the Townsend-George model performs better than common wind turbine wake models, notably in predicting the evolution of the centreline velocity deficit. A similar conclusion was drawn for different operating tip-speed ratios, \emph{i.e.} different thrust coefficients \cite{Neunaber2022_b}. It has been shown that the classical bluff body wake models perform better due to the presence of a virtual origin $x_0$ in the scaling, and implementing this parameter in the classical wake models significantly improves their performance \cite{Neunaber2022,Neunaber2024}. In the present section, we will investigate the applicability of the Townsend-George theory to disc-generated wakes developing in different turbulence “flavours”, as well as the effects of the porosity/thrust coefficient and FST on the virtual origin, which has been shown to be the leading parameter for wind turbine wake modelling \cite{Neunaber2024} .

\begin{figure}[tp]
    \centering
     \begin{minipage}[c]{0.7\textwidth}
          \begin{ruledtabular}
         \captionof{table}{Theoretical axial scaling laws for the streamwise evolution the maximum velocity deficit in turbulent axisymmetric wakes. \label{Tab2}}
    \begin{tabular}{ccc}
     Equilibrium scaling (\emph{e.g.} \cite{George1989}) & $\Delta U_{max} \sim (x-x_0)^{-2/3}$ \\
    Square-root non-equilibrium scaling (\emph{e.g.} \cite{Vassilicos2015}) & $\Delta U_{max} \sim (x-x_0)^{-1}$ \\
    Linear non-equilibrium scaling scaling (\emph{e.g.} \cite{Stein2019})  & $\Delta U_{max} \sim (x-x_0)^{-2}$ 
    \end{tabular}
    \label{tab:scaling laws}
    \end{ruledtabular}
    \end{minipage}
\end{figure}

\subsection{Verification of the requirements}

Neunaber~\emph{et al.}~\cite{Neunaber2022,Neunaber2021_Townsend} provided a practical and comprehensive summary of the prerequisites necessary for applying the Townsend-George theory to wind turbine wakes. We will adopt the same methodology to ascertain whether porous and solid discs-generated wakes meet these requirements. TABLE~\ref{tab:summary} summarises the investigation process and indicates the conditions under which the different criteria allowing the application of the Townsend-George theory are fulfilled, or not, for all $\{\textrm{disc}/\textrm{FST case}\}$ pairs. 

The first requirement to apply the Townsend–George theory is fully developed, decaying turbulence, which is indicated by a decreasing centreline turbulence intensity $TI_{y=0}$. TABLE~\ref{tab:summary} displays the streamwise position $x/D$ at which $TI_{y=0}$ begins to decay for all $\{\textrm{disc}/\textrm{FST case}\}$ configurations, as also depicted in FIG~\ref{Fig:TIcentreline}. 

\begin{figure}[btp]
    \centering
     \begin{minipage}[t]{0.49\textwidth}
        \includegraphics[width=\textwidth]{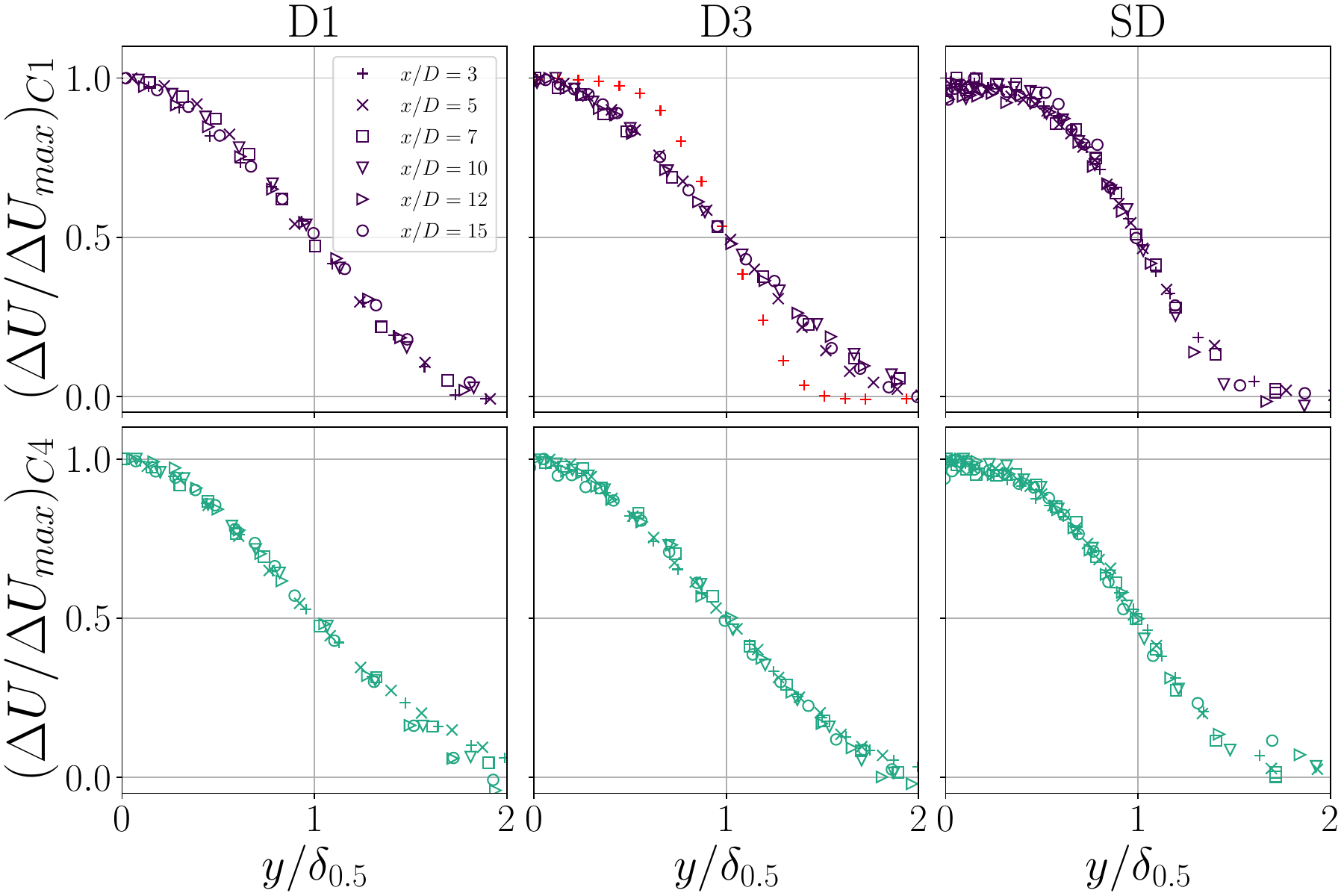} 
        \caption{Evolution of the normalised velocity deficit $\Delta U/\Delta U_{max}$ with the normalised horizontal position $y/ \delta_{0.5}$ for D1, D3 and SD in Case 1 (top) and Case 4 (bottom). The red crosses \textcolor{red}{+} show that self similarity is not achieved at $x/D=3$ for the configuration $\{\textrm{D3}/\textrm{Case 1}\}$. }
        \label{Fig:SelSIm}
    \end{minipage}
    \hfill
    \begin{minipage}[t]{0.47\textwidth}
        \centering
                \includegraphics[width=\textwidth]{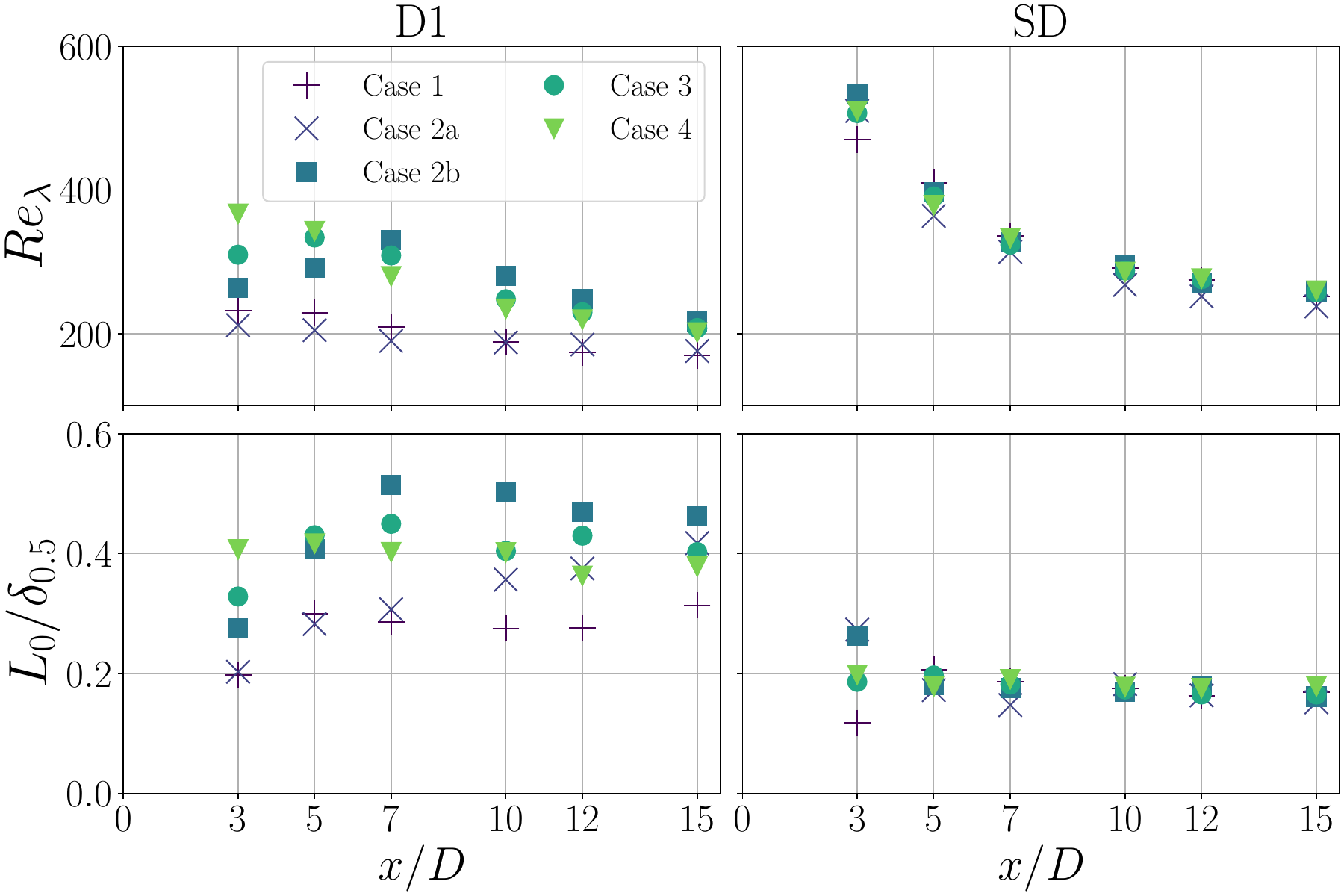} 
        \caption{Evolution of (top) the Taylor Reynolds number $Re_{\lambda}$ and (bottom) the normalised integral length scale $L_0/\delta_{0.5}$ with increasing distance $x/D$ downstream of D1 (left) and SD (right).}
        \label{Fig:Taylor}    
    \end{minipage}  
\end{figure}

Secondly, the mean streamwise velocity and turbulence quantities (Reynolds stresses, turbulent kinetic energy, dissipation rate of turbulent kinetic energy) have to be self-similar to apply the Townsend-George theory \cite{Nedic2013,Dairay2015}. As we conducted 1D hot-wire anemometry in the present study, our examination of self-similarity is limited to the mean velocity profile. It is important to keep in mind that the streamwise location $x/D$ in the wake where the mean velocity deficit achieves self-similarity does not coincide with the location where higher-order turbulent quantities reach this state. Typically, the latter is attained further downstream in the wake. However, as is common practice for wind turbine wakes \cite{Neunaber2022,Neunaber2021_Townsend}, we will focus solely on the self-similarity of the mean velocity deficit. To verify this requirement, we plotted the normalised mean velocity deficit $\Delta U/\Delta U_{max}$ against the normalised radial coordinate $y/\delta_{0.5}$ for all wakes. The profiles for three different discs and two FST cases are displayed in FIG.~\ref{Fig:SelSIm}. The self-similar behaviour is checked when the curves collapse. One can observe that for the configuration $\{\textrm{D3}/\textrm{Case 1}\}$, self-similarity is not achieved at $x/D = 3$ (marked by red crosses \textcolor{red}{+}), whereas for the other profiles displayed, all curves collapsed for $x/D \geq 3$. For all $\{\textrm{disc}/\textrm{FST case}\}$ pairs, self-similarity of ($\Delta U/\Delta U_{max}$) is reached between $x/D =3$ and $x/D =5$ (the locations are reported in the second column of TABLE~\ref{tab:summary}).

Thirdly, the flow must be turbulent. As suggested by Neunaber~\emph{et al.}~\cite{Neunaber2022}, one method to verify this criterion is (i) to check the presence in the PSD of fluctuating velocity ($E_{u'u'}$) of an inertial sub range that decays according to a $-5/3$ law, \emph{i.e.} $E_{u'u'} \varpropto f^{-5/3}$, and (ii) to ensure that the Taylor Reynolds number $Re_{\lambda}$ is sufficiently large, $Re_{\lambda}\geq 200$. All wakes fulfilled the (i) criterion (FIG.~\ref{Fig:Spectra} displays several examples of PSDs with a $f^{-5/3}$ law overlaid). The Taylor Reynolds numbers along the centreline of all wakes have been calculated in the same manner as Neunaber \emph{et al.} in \cite{Neunaber2022}, and the streamwise evolutions of $Re_{\lambda}$ for two distinct discs and FST cases are depicted in FIG.~\ref{Fig:Taylor}. Requirement (ii) of the Townsend-George theory is satisfied in all wakes ($Re_{\lambda} \geq 200$ ), except in the far wakes of D1 in Cases 1 and 2a (see third and fourth columns of TABLE~\ref{tab:summary})

Fourthly, the mean velocity profiles and turbulence quantities have to be axisymmetric. Although the tower disrupts the wake axisymmetry, similar to what is observed in wind turbine wakes \cite{Santoni2017,Pierella2017}, a closer examination of the velocity deficit and $TI$ profiles in large scale (FIG.~\ref{Fig_AppendixA} \& FIG.~\ref{Fig_AppendixB} in Appendix~\ref{AppendixA}) reveals that this criterion is satisfactorily fulfilled for all wakes.

The final requirement is that the flow has to be in a streamwise range for which the longitudinal integral length scale is proportional to the wake width, \emph{i.e.} $L \varpropto \Delta \delta_{0.5}$. We calculated the centreline integral length scale $L$ for all wakes using the same approach as for determining the integral length scale of the turbulent ambient flow $L_0$. $L/\Delta \delta_{0.5}$ is plotted against $x/D$ for two different discs and FST cases in FIG.~\ref{Fig:Taylor}. In certain wakes, the condition $L/\Delta \delta_{0.5}$ being approximately constant is met, in particular when the ambient flow is highly turbulent (Case 4). However, for low turbulent ambient cases (Group 1 cases), $L/\Delta \delta_{0.5}$ tends to increase with the streamwise distance. The last column of TABLE~\ref{tab:summary} indicates the $\{\textrm{disc}/\textrm{FST case}\}$ pairs for which this condition is satisfactorily met.

It can be noted that most of the prerequisites for applying the Townsend-George theory are satisfactorily fulfilled for all wakes within either $x/D \geq 3$ or $x/D \geq 5$ (TABLE~\ref{tab:summary}). The condition of $L/\Delta \delta_{0.5}$ being approximately constant is the most restrictive condition for applying the theory, and in some wakes this requirement is not fulfilled. The Townsend-George theory requirements are then only partially fulfilled. It is important to note that, in the Townsend-George theory, the turbulent axisymmetric wake develops within a non-turbulent ambient. However, in the present study, the ambient flow is turbulent in many cases, which could potentially impact the feasibility of meeting certain criteria, such as self-similarity \cite{Rind2012}. We will now proceed by fitting the various scaling laws (TABLE~\ref{tab:scaling laws}) to all wakes from two different points $x_{fit} = 3D$ and $x_{fit} = 5D$, bearing in mind that for certain wakes the Townsend-George theory requirements are only partially fulfilled. By fitting experimental data from two different origins, we aim to explore the repercussions of neglecting preliminary checks on Townsend-George theory requirements before employing classical wake laws for bluff bodies. Additionally, given the lack of consensus on the physical starting position of the far wake and the variability in reported fitting origin positions in the literature, we seek to determine whether the choice of fitting starting position influences the outcomes.

\begin{table}[]
    \centering
    \begin{ruledtabular}

    \begin{tabular}{c| c | ccc c c c }
      \multirow{2}*{Disc} & \multirow{2}*{FST case} & \textit{Requirement 1} &  \textit{Requirement 2} &  \textit{Requirement 3(i)} &  \textit{Requirement 3(ii)} & \textit{Requirement 4} \\
        &  & Decaying $TI_{y=0}$ & Self-similarity & $E_{u' u'} \varpropto f^{-5/3}$ & $Re_{\lambda} > 200 $ & $L_{y=0} \varpropto \delta_{0.5}$ \\
        
        \hline
        \multirow{5}*{D1} & Case 1   & $x/D \geq 3$  & $x/D \geq 3$    & \checkmark     &  \textcolor{red}{(\checkmark)} & \textcolor{red}{$x/D \geq 5$} \\
         & Case 2a  & $x/D \geq 3$  & \textcolor{red}{$x/D \geq 5$}    &   \checkmark &  \textcolor{red}{(\checkmark)}  &    \textcolor{red}{$\times$}   \\
         & Case 2b  & $x/D \geq 3$  & $x/D \geq 3$                     &  \checkmark  &  \checkmark & \textcolor{red}{$x/D \geq 7$}  \\
         & Case 3   & $x/D \geq 3$  & $x/D \geq 3$                     &  \checkmark  &  \checkmark  & \textcolor{red}{$x/D \geq 5$} \\
         & Case 4   & $x/D \geq 3$  & $x/D \geq 3$                     &  \checkmark  &  \checkmark  & \checkmark \\
        \hline

\multirow{5}*{D2} & Case 1   & \textcolor{red}{$x/D \geq 5$}   & \textcolor{red}{$x/D \geq 5$}    & \checkmark     &   \checkmark & \textcolor{red}{$\times$} \\
         & Case 2a  & \textcolor{red}{$x/D \geq 5$}  & \textcolor{red}{$x/D \geq 5$}    &   \checkmark &   \checkmark &    \textcolor{red}{$\times$}   \\
         & Case 2b  & \textcolor{red}{$x/D \geq 5$}  & $x/D \geq 3$  &  \checkmark  &  \checkmark & \textcolor{red}{$x/D \geq 5$}  \\
         & Case 3   & $x/D \geq 3$  & $x/D \geq 3$                     &  \checkmark  &  \checkmark  & \textcolor{red}{$x/D \geq 5$} \\
         & Case 4   & $x/D \geq 3$  & $x/D \geq 3$                     &  \checkmark  &  \checkmark  & \checkmark \\
        \hline
        
\multirow{5}*{D3} & Case 1   & \textcolor{red}{$x/D \geq 5$}   & \textcolor{red}{$x/D \geq 5$}    & \checkmark     &   \checkmark & \textcolor{red}{$\times$} \\
         & Case 2a  & \textcolor{red}{$x/D \geq 5$}  & \textcolor{red}{$x/D \geq 5$}    &   \checkmark &   \checkmark &    \textcolor{red}{$\times$}   \\
         & Case 2b  & \textcolor{red}{$x/D \geq 5$}  & $x/D \geq 3$  &  \checkmark  &  \checkmark & \textcolor{red}{$x/D \geq 5$}  \\
         & Case 3   & $x/D \geq 3$  & $x/D \geq 3$                     &  \checkmark  &  \checkmark  & \textcolor{red}{$x/D \geq 5$} \\
         & Case 4   & $x/D \geq 3$  & $x/D \geq 3$                     &  \checkmark  &  \checkmark  & \checkmark \\
         \hline
\multirow{5}*{D4} & Case 1   & $x/D \geq 3$    & \textcolor{red}{$x/D \geq 5$}    & \checkmark     &   \checkmark & \textcolor{red}{$\times$} \\
         & Case 2a  & $x/D \geq 3$  & \textcolor{red}{$x/D \geq 5$}    &   \checkmark &   \checkmark &    \textcolor{red}{$\times$}   \\
         & Case 2b  & $x/D \geq 3$  & $x/D \geq 3$  &  \checkmark  &  \checkmark & \textcolor{red}{$x/D \geq 5$}  \\
         & Case 3   & $x/D \geq 3$  & $x/D \geq 3$                     &  \checkmark  &  \checkmark  & \textcolor{red}{$x/D \geq 5$} \\
         & Case 4   & $x/D \geq 3$  & $x/D \geq 3$                     &  \checkmark  &  \checkmark  & \textcolor{red}{$x/D \geq 5$} \\    
         \hline
         \multirow{5}*{SD} & Case 1   & $x/D \geq 3$  & $x/D \geq 3$     & \checkmark     &   \checkmark & \textcolor{red}{$x/D \geq 5$}\\
                           & Case 2a  & $x/D \geq 3$  & $x/D \geq 3$     &   \checkmark &   \checkmark &    \textcolor{red}{$x/D \geq 5$}   \\
                           & Case 2b  & $x/D \geq 3$  & $x/D \geq 3$     &  \checkmark  &  \checkmark & \textcolor{red}{$x/D \geq 5$}  \\
                           & Case 3   & $x/D \geq 3$  & $x/D \geq 3$     &  \checkmark  &  \checkmark  &  \checkmark \\
                           & Case 4   & $x/D \geq 3$  & $x/D \geq 3$     &  \checkmark  &  \checkmark  & \checkmark \\

    \end{tabular}
    \caption{Summary of conditions at which the Townsend-George theory requirements are fulfilled in wakes of all $\{\textrm{disc}/\textrm{FST case}\}$ configurations. \checkmark denotes that the requirement is fulfilled, (\checkmark) indicates the requirement is partially fulfilled, and ($\times$) that the requirement is not fulfilled. Requirements not met or fulfilled further downstream from the closest measurement location are highlighted in \textcolor{red}{red}. }
    \label{tab:summary}
    \end{ruledtabular}

\end{table}

\subsection{Results}

\subsubsection{Centreline Velocity Deficit Fitting}

\begin{figure}[hbtp]
    \centering
     \begin{minipage}[c]{0.47\textwidth}
        \includegraphics[width=\textwidth]{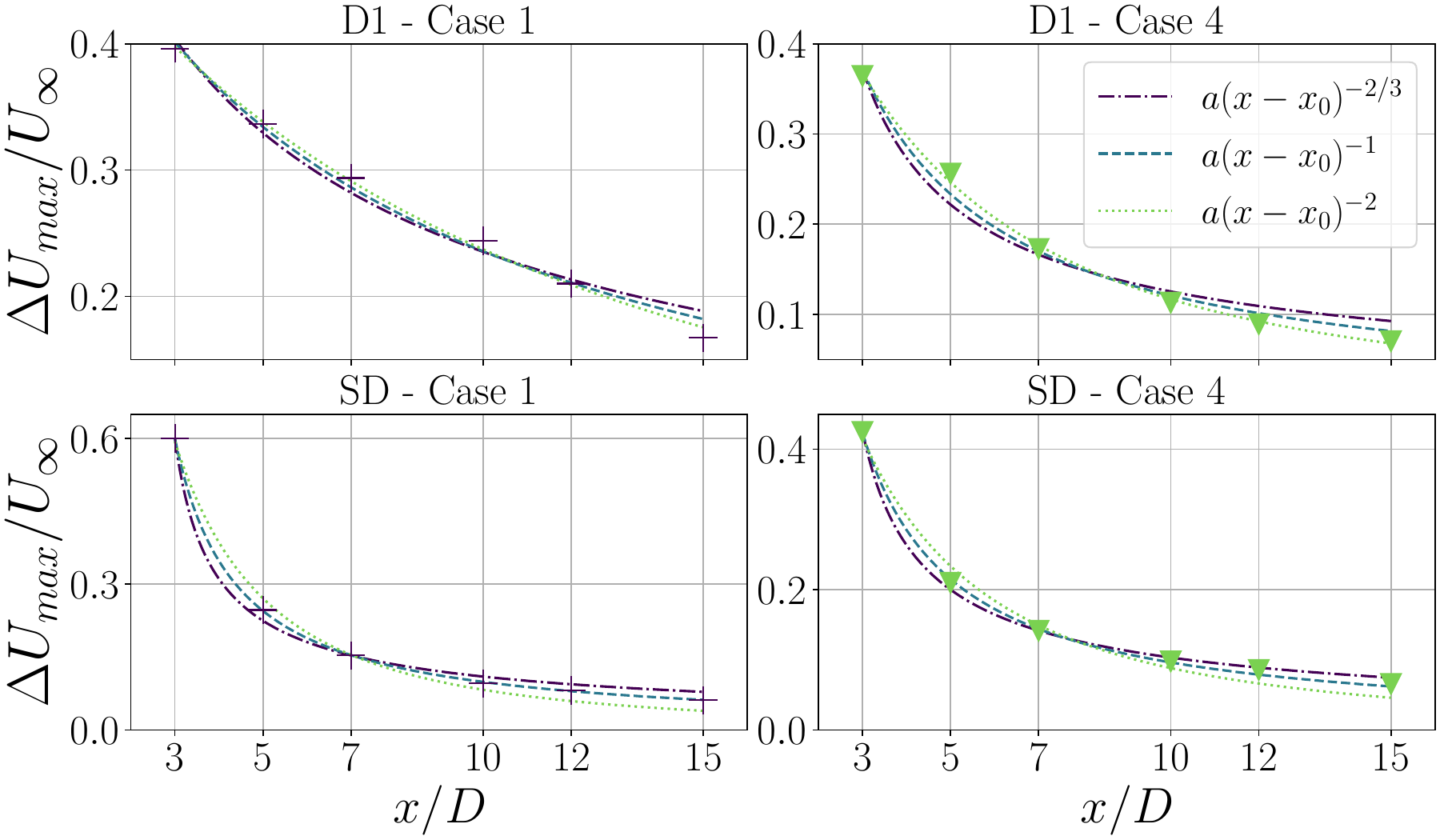} 
        \caption{Power law fits of various models (TABLE.~\ref{tab:scaling laws}) for the streamwise evolution of $\Delta U_{max}/U_{\infty}$ for D1 (top) and SD (bottom) in Case 1 (left) and Case 4 (right). The experimental data have been fitted from $x/D=3$.}        \label{Fig:Fit_Velo_Def}
    \end{minipage}
    \hfill
    \begin{minipage}[c]{0.47\textwidth}
        \centering            
        \includegraphics[width=\textwidth]{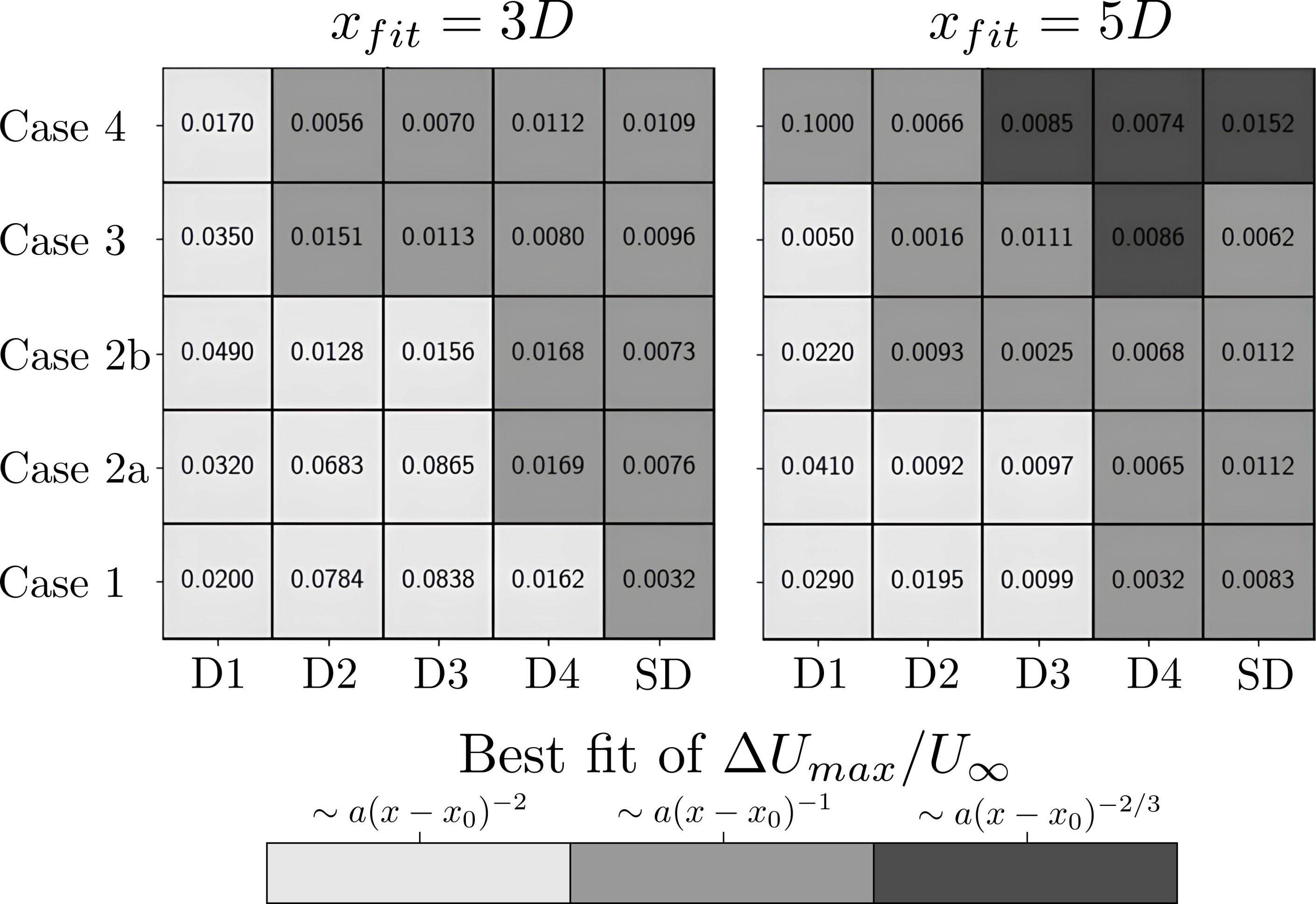} 
                \caption{Scaling laws providing the most accurate fit for the streamwise evolution of $\Delta U_{max}/U_{\infty}$ for all $\{\textrm{disc}/\textrm{FST case}\}$ configurations. Each scaling law is represented by a grayscale, with the color in each box corresponding to the fit that minimises the normalised root-mean-square errors (displayed at the centre of each box). The various models (TABLE.~\ref{tab:scaling laws}) have been fitted from $x_{fit} = 3D$ (left) and $x_{fit}=5D$ (right). }
                \label{Fig:Best_Fit}  
        \end{minipage} 
\end{figure}

After verifying all requirements, we fitted the three theoretical axial scaling laws (TABLE~\ref{Tab2}) to the entire dataset of streamwise evolution of $\Delta U_{max}/U_{\infty}$ (FIG.~\ref{Fig:MaxVeloDef}), with the fits starting from two distinct positions $x_{fit}=3D$ and $x_{fit}=5D$. FIG.~\ref{Fig:Fit_Velo_Def} shows the streamwise evolution of ($\Delta U_{max}/U_{\infty}$) for D1 and SD in two FST “flavours”, with the power-law fits performed from $x_{fit} = 3D$. The normalised root-mean-square errors (NRMSE) were computed to assess the quality of each regression, as detailed in FIG~\ref{Fig:Best_Fit}. In this figure, each scaling law is assigned a greyscale, and the colour of each box corresponds to the “best fit", \emph{i.e.} the fit with the lowest NRMSE. The NRMSE values are displayed at the centre of each box. 

Firstly, it is noteworthy that despite certain requirements of the Townsend-George theory not being fulfilled, the different scaling laws fit the velocity deficits quite well, with relatively low NRMSE values. Interestingly, one can observe a progressive modification of the scaling law that minimises the NRMSE, depending on the disc porosity, the FST conditions, and the origin of the fit. As the background turbulence increases, along with the solidity or the starting position of the fit, there is a gradual shift in the optimal fitting model -- from a $\Delta U_{max} \varpropto x^{-2}$ power law at low thrust coefficient and under low ambient turbulence conditions, to a $\Delta U_{max} \varpropto x^{-1}$ power law (when $x_{fit} = 3D$) and ultimately a $\Delta U_{max} \varpropto x^{-2/3}$ power law (when $x_{fit} = 5D$) at high thrust coefficients and under high ambient turbulence levels. The main discrepancies in NRMSE between the three fits occur with the most porous disc (D1) and in Case 1 with no FST in the ambient flow, likely because the wake required more distance to develop fully. For the solid disc, all three fits satisfactorily describe the evolution of $\Delta U_{max}/\Delta U$, although we can observe a tendency of the exponent to increase as either $TI_{\infty}$ or $L_0$ increases. 

Lingkan \& Buxton~\cite{Lingkan2023} reported that the most accurate scaling law for predicting the evolution of $\Delta U_{max}$ downstream of a wire mesh porous disc with uniform porosity of $\beta = 0.45$, was the $-2$ power law. In their study, the ambient flow was non-turbulent, and the fits were conducted from $x_{fit} = 3D$. We note a similar outcome with all four porous discs in Case 1 (no FST) when fitting from $x_{fit} = 3D$. Chongsiripinyo \& Sarkar's~ \cite{Chongsiripinyo2020} Large Eddy Simulation (LES) study on the decay of turbulent wakes downstream of a solid disc has also revealed the existence of two distinct wake regions characterised by different power law exponents governing the decay of $\Delta U_{max}$. Between $10 < x/D < 65$, the authors observed an evolution following $\Delta U_{max} \varpropto x^{-1}$, while further downstream, within $65 < x/D < 125$, they identified $\Delta U_{max} \varpropto x^{-2/3}$. In the present study, the measurement range is limited to 15 diameters. Hence, it is unsurprising that for the solid disc, the best scaling law in the absence of FST (Case 1) does not change from $\Delta U_{max} \varpropto x^{-1}$ to a $\Delta U_{max} \varpropto x^{-2/3}$ by changing the initial fit position from $x_{fit}=3D$ to $x_{fit}=5D$. However, in the presence of FST (Case 4), the change in the power law observed by changing the starting position of the fit might indicate that we capture the transition region that separates these two stages. This transition occurs closer to the disc in presence of FST, which is also consistent with the faster wake recovery.

\subsection{Effect of FST and porosity on the virtual origin $x_0$}

While a virtual origin $x_0$ emerges naturally in the Townsend-George theory, conventional “engineering" wake models typically do not include this parameter in their original formulations. Using datasets from both lab-scale and full-scale wind turbines, Neunaber, Hölling \& Obligado~\cite{Neunaber2024} observed that incorporating a virtual origin into “engineering" wake models significantly enhances their accuracy in predicting the evolution of the centreline velocity deficit. Therefore, the authors postulated that the most important parameter for wind turbine wake modelling is the virtual origin $x_0$, as all models achieve better fits simply by adding this virtual origin. They finally underscore the necessity for further investigation to clarify the dependence of the virtual origin on both freestream turbulence conditions and thrust coefficient. We aim to address this need through the experiments conducted in the present study, which enables us to explore an extensive parameter space $\{C_T; L_0; TI_{\infty}\}$.


\begin{figure}[hbtp]
    \centering
     \begin{minipage}[c]{0.47\textwidth}
        \includegraphics[width=\textwidth]{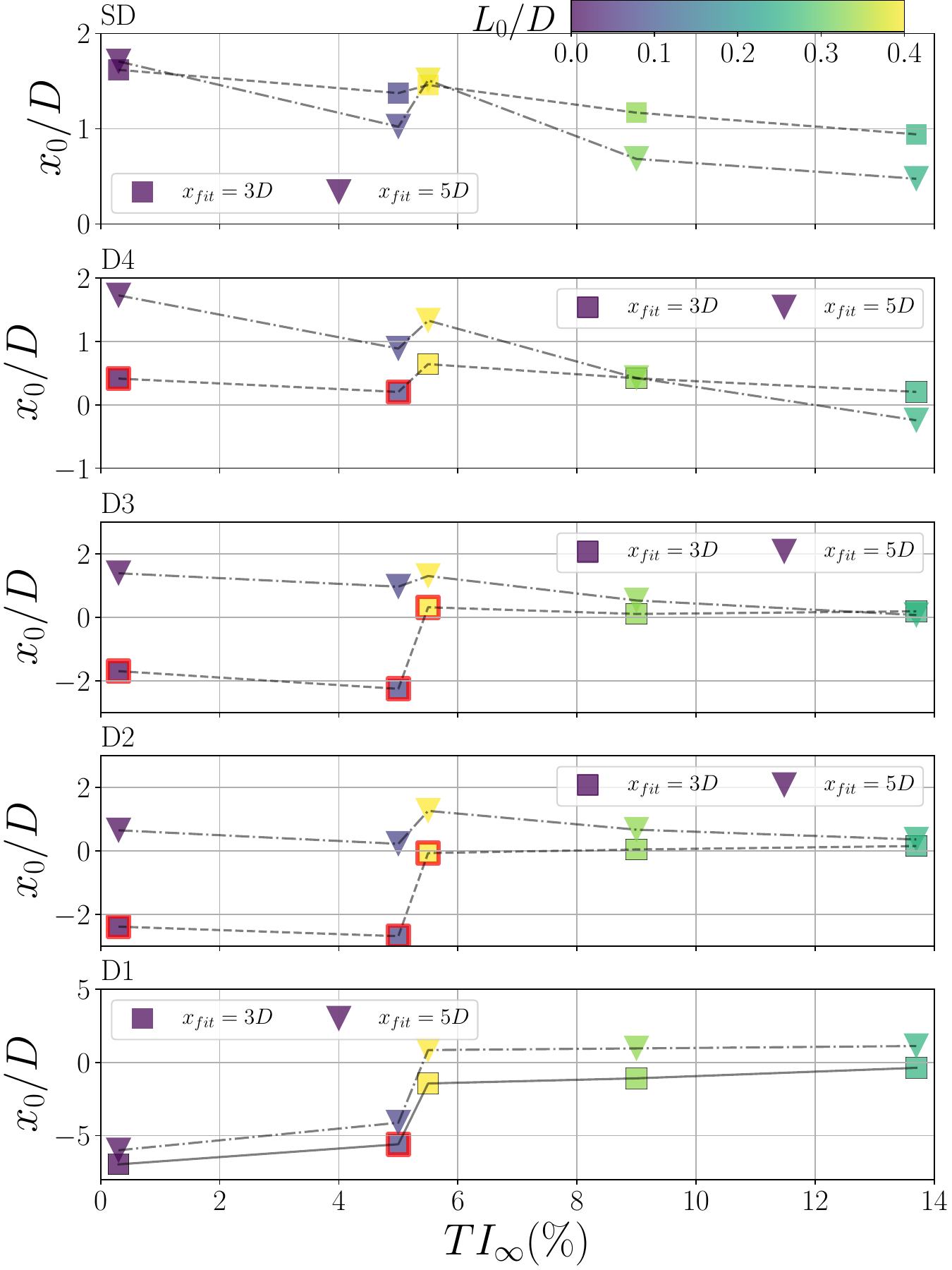} 
        \caption{Virtual origin $x_0/D$ for all $\{\textrm{disc}/\textrm{FST case}\}$ pairs as a function of the freestream turbulence intensity $TI_{\infty}$ (horizontal axis) and integral length scale $L_0/D$ (colour axis). Virtual origins are obtained through fitting the centreline velocity deficit evolution with the following power law $\Delta U_{max} = a(x-x_0)^{-1}$, with fits starting from $x_{fit} = 3D$ (squares) and $x_{fit} = 5D$ (triangles).  Squares outlined in red indicate cases where either self-similarity or decaying $TI$ requirements are not met.}
        \label{Fig:virtual_orign_TI}
    \end{minipage}
    \hfill
    \begin{minipage}[c]{0.47\textwidth}
        \centering
         \begin{minipage}[c]{\textwidth}                
        \includegraphics[width=0.9\textwidth]{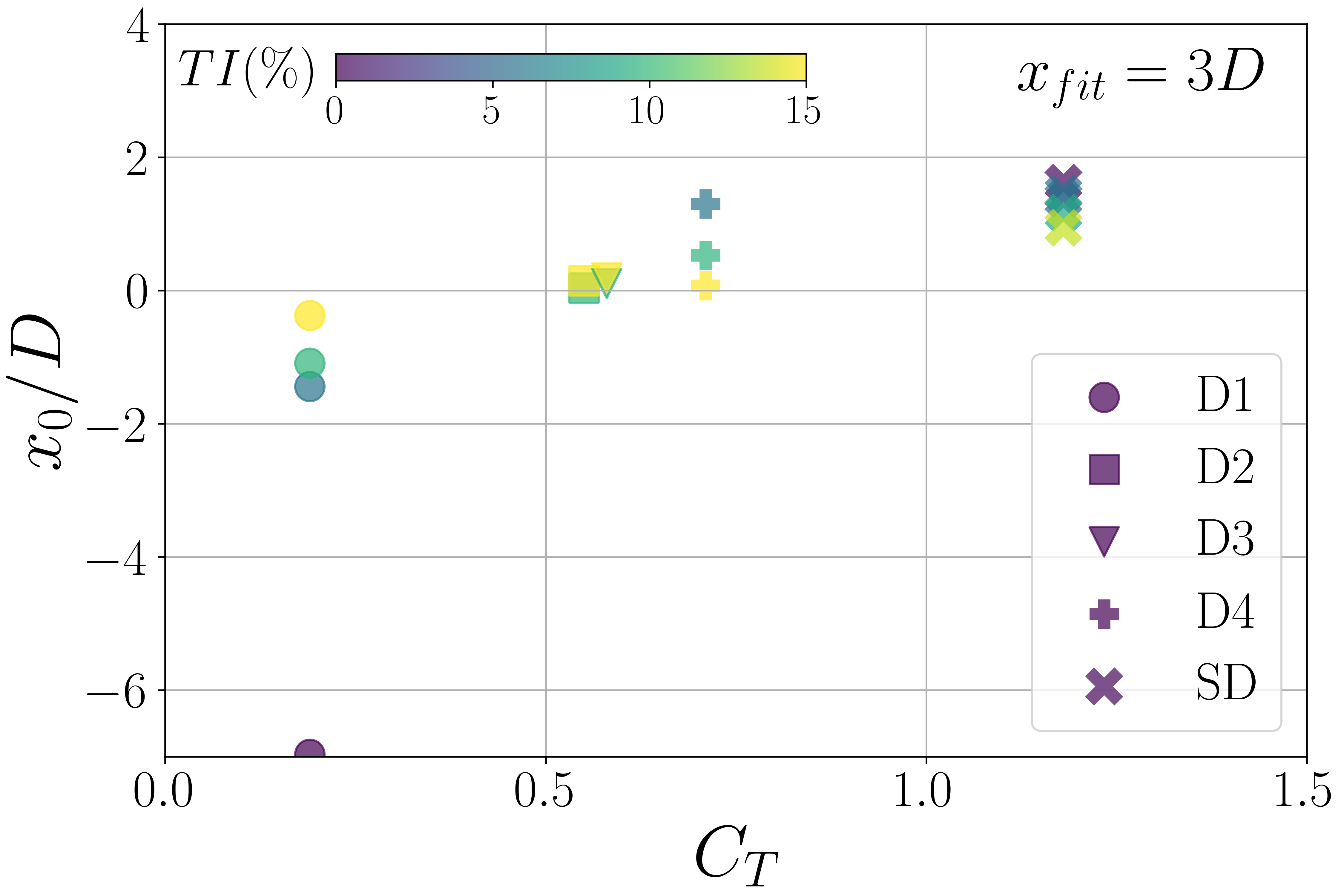} 
                \subcaption{}
                \label{Fig:virtual_orign_Ct3D}  
             \end{minipage} 
                 \\
         \begin{minipage}[c]{\textwidth}    
            \includegraphics[width=0.9\textwidth]{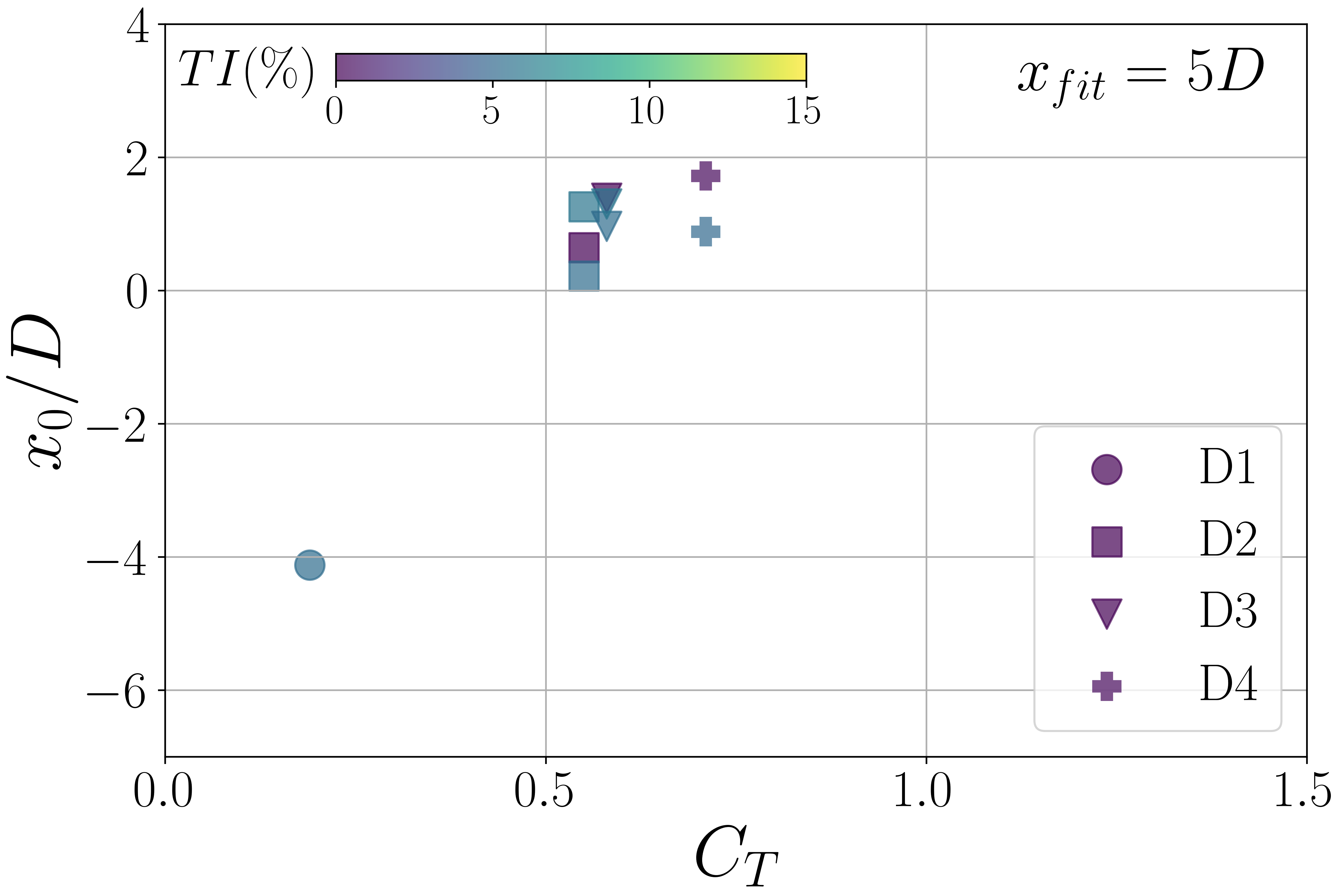} 
            \subcaption{}
            \label{Fig:virtual_orign_Ct5D}  
         \end{minipage}
        \caption{Virtual origin $x_0/D$ for all $\{\textrm{disc}/\textrm{FST case}\}$ pairs as a function of the thrust coefficient $C_T$ (horizontal axis) and freestream turbulence intensity $TI_{\infty}$ (colour axis). Results are presented in two graphs. Configurations meeting requirements 1, 2 and 3 of the Townsend-George theory (TABLE~{\ref{tab:summary}}) for $x/D \geq 3$ are shown in (\subref{Fig:virtual_orign_Ct3D}), and for $x/D \geq 5$ in (\subref{Fig:virtual_orign_Ct5D}).
        }
            \label{Fig:virtual_orign_Ct} 
    \end{minipage} 
\end{figure}

It is important to emphasise that comparing the virtual origins $x_0$ of different $\{\textrm{disc}/\textrm{FST case}\}$ pairs is only appropriate when the evolution of $\Delta U_{max}$ has been fitted using the same scaling laws and starting points $x_{fit}$. Therefore, in FIG.~\ref{Fig:virtual_orign_TI}, we report the virtual origins for all wakes obtained by fitting the respective centreline velocity deficit evolution with the same power law $\Delta U_{max} \varpropto x^{-1}$, starting either form $x_{fit} = 3D$ (square symbols) or $x_{fit} = 5D$ (triangle symbols). We verified that fitting with the two other scaling laws does not change the observed trends concerning the influence of the porosity/thrust coefficient or FST on $x_0$. It solely modifies the magnitude of the fitting coefficients. Moreover, beginning fits from different streamwise positions $x_{fit}$ changes the coefficients of the fits. However, as observed in TABLE~\ref{tab:summary}, the Townsend-George theory requirements of self similarity and decaying turbulence are not fulfilled at the same positions $x/D$. For the solid disc, these two requirements are fulfilled for all FST cases at $x/D = 3$, but for all porous discs, these requirements are achieved either at $x/D = 3$ or $x/D = 5$. As a relevant comparison of the virtual origins $x_0$ for the various wakes can only be conducted if the fits start from the same position $x_{fit}$, we plot in FIG.~\ref{Fig:virtual_orign_TI} the virtual origins obtained from fitting form the two origins $x_{fit} = 3D$ and $x_{fit} = 5D$ for all FST cases. The squares with red edges indicates that either the condition of self-similarity or decaying $TI$ is not fulfilled for the corresponding $\{\textrm{disc}/\textrm{FST case}\}$ pair.

The range of virtual origins $x_0/D$ obtained for the different disc-generated wakes closely aligns with the values reported for wakes generated by both lab-scale and full-scale wind turbines operating under various FST conditions \cite{Neunaber2024}, indicating that porous discs can also faithfully replicate this quantity. For all discs, we can identify that increasing the turbulence intensity of the background $TI_{\infty}$ results in a convergence of the virtual origin towards $x_0/D = 0$, regardless of whether $x_{fit} = 3D$ or $x_{fit} = 5D$ (FIG.~\ref{Fig:virtual_orign_TI}). A similar effect of the FST on the virtual origin in wind turbines-generated wakes has also been observed by Neunaber, Hölling \& Obligado \cite{Neunaber2024}, and by Gambuzza \& Ganapathisubramani \cite{Gambuzza2022}. For all discs, the virtual origin is farthest from $x_0 = 0$ in Case 1 (no FST), which can be related to an extended near wake region \cite{Neunaber2024}. This is consistent with the observation of faster wake recovery in the presence of FST, along with the observed initial wake width reduction for the porous discs in the absence of FST suggesting a less developed wake state.

For SD and D1, fitting from either $x_{fit} = 3D$ or $x_{fit} = 5D$ does not alter the evolution of $x_0$ with increasing ambient turbulence intensity $TI_{\infty}$ or integral length scale $L_0$. However, for SD $x_0$ decreases when increasing $TI_{\infty}$ whereas it is the opposite for D1. For all discs, we observe a clear effect of the integral length scale, with $x_0$ increasing as $L_0$ increases (when comparing cases 2a and 2b). For the three intermediate porosity discs D2, D3 and D4, the evolution of $x_0$ with $TI_{\infty}$ varies depending on the origin of the fit $x_{fit}$, even though, for both $x_{fit}$, the virtual origins tend towards $x_0/D = 0$ as $TI_{\infty}$ increases. Interestingly, the main discrepancies in $x_0$ between the fits from $x_{fit}=3D$ and $x_{fit}=5D$ occur in the FST cases where the criteria of self-similarity and decaying turbulence are not fulfilled at $x_{fit}=3D$ (red-bordered squares). Without considering these two requirements and consequently looking at the fits with $x_{fit}=3D$, we observe the same trend as for D1, namely an increase in $x_0$ as $TI_{\infty}$ increases. However, by starting the fit only after reaching a point where the requirements of self similarity and decaying turbulence are met for all FST cases ($x_{fit} = 5D$), one can observe a behaviour akin to that of the solid disc: a reduction in the virtual origin as $TI_{\infty}$ increases. This result highlights the importance of checking the requirements of the Townsend-George theory before fitting scaling laws to experimental data.

To compare the discs and examine the influence of porosity/thrust coefficient on the virtual origin, we selected in FIG.~\ref{Fig:virtual_orign_Ct3D} only the $\{\textrm{disc}/\textrm{FST case}\}$ configurations where both self-similarity and decaying turbulence conditions are satisfied at $x/D=3$, and in FIG.~\ref{Fig:virtual_orign_Ct5D}, configurations where these conditions are met at $x/D = 5$. In both figures, there is a clear increase of the virtual origin as the thrust coefficient increases, \emph{i.e.} as the porosity decreases. This is consistent with observations indicating that the strength and size of the recirculation bubble, which are essentially confined within the immediate wake and therefore serve as an indicator of the near-wake length downstream of a disc, initially diminish and subsequently become suppressed with increasing body porosity \cite{Steiros2021}. 

To summarise, after reaching the requirements for self-similarity and decaying turbulence, the virtual origin decreases as the turbulence intensity of the background increases for D2, D3, D4 and SD, whereas it is the opposite for D1. It highlights that FST effects are related to the disc porosity and thrust coefficient, which has also been shown with previous wake characteristics. Moreover, increasing the integral length scale of ambient turbulence or the thrust coefficient increases $x_0$. Finally, we emphasize the importance of verifying the Townsend-George theory requirements before applying the classical scaling laws. Even though not all requirements are met (especially $L \varpropto \Delta \delta_{0.5}$), ensuring the self-similarity of the mean velocity profiles and the decaying turbulence intensity along the centreline appears to be indispensable for drawing meaningful conclusions. Further research should focus on gaining a deeper understanding of the relationship between the diverse flow phenomena in the near wake and the virtual origin. For instance, the existence of very low and negative values of $x_0/D$ for the most porous disc D1 in low FST conditions (which are also observed in wind turbine wakes \cite{Neunaber2024}) raises some interesting questions regarding the physics that control this value.

\section{Conclusions}

A study on the development of wakes generated by discs of varying porosity within a turbulent background, characterised by different levels of turbulence intensity and integral length scale has been presented. The parameter space explored encompasses 5 levels of porosity/thrust coefficient and 5 freestream turbulence “flavours”, resulting in a total of 25 $\{\textrm{disc}/\textrm{FST case}\}$ configurations. The porosity of the discs was extensively varied, ranging from $\beta = 0$ to $\beta = 0.6$, in order to explore a broad range of thrust coefficients, spanning from $C_T = 1.18$ to $C_T = 0.19$. The turbulence intensity $TI_{\infty}$, and integral length scale $L_0$ of the ambient flow were independently varied to examine their individual effects on the evolution of the wakes produced by the discs.

The shape of the velocity deficit profiles and turbulence intensity profiles in the near wake, as well as their evolution with increasing streamwise distance, strongly vary depending on $TI_{\infty}$, $L_0$, and $\beta$. For all discs, FST accelerates the transition of the wake to a developed state characterised by self-similarity of the velocity deficit profiles and decreasing turbulence intensity. The velocity deficit recovery and homogenisation of the turbulence intensity in the wake $TI$ with that of the ambient $TI_{\infty}$ are accelerated in the wake of low-porosity discs, likely due to the high levels of turbulence intensity added to the flow. This, in turn, limits the additional impact of the ambient turbulence intensity on the velocity deficit recovery and decay of $TI$, or at least restricts it to a smaller distance from the disc. For low porosity discs, since the turbulence intensity added to the flow is low, the ambient turbulence has a stronger influence on the wake, which persists over a longer distance.


The impact of FST on the wake width, wake growth rate, and entrainment rate also depends on the disc porosity and thrust coefficient. For low ambient turbulence conditions (Group 1 FST cases), an initial wake contraction region is observed downstream of the porous discs, and its length increases as the porosity increases. This initial wake contraction coincides with the presence of a low-frequency peak in the near wake spectra of the velocity fluctuations. The large-scale coherent structures shed by the disc might act as a shield against mixing in the near wake, thereby preventing the near wake growth. We found that FST suppresses, or at least weakens, the energy content of these large-scale coherent structures, thus limiting their impact on the mixing process to a shorter distance downstream of the discs. This mirrors observations made with wind turbine models and wings, indicating an earlier onset of tip-vortex breakdown when subjected to FST \cite{Odemark2013,Kamal2020}. Consequently, close to the discs, the wake is wider and the mass-flux is larger in the presence of FST. Further downstream in the wake, the influence of this structure on wake expansion is attenuated. We observed that, at a sufficient distance downstream, the wake growth rates and entrainment rates are higher when the background is non-turbulent for all discs, except for the most porous disc D1, for which both the wake growth rate and entrainment rates increase with FST. As the porosity of the disc decreases, the influence of FST on the entrainment behavior gradually aligns with those observed with solid bluff bodies; however, a singular effect of FST is observed at high porosity. This demonstrates that FST effects on porous disc-generated wakes depend strongly on the disc porosity and thrust coefficient. 

The applicability of the Townsend–George theory to disc-generated wakes for the different incoming flows was then examined. It has been found that, akin to wind turbine wakes \cite{Neunaber2021_Townsend,Neunaber2022}, the requirements of this theory are either satisfied or partially fulfilled in the different wakes depending on $\{ TI_{\infty}; L_0 ; \beta \}$. Classical theoretical scaling laws were fitted to the centreline velocity deficit of all wakes, revealing a clear effect on the power exponent of the best-fitting law depending on the starting position of the fit, the porosity, and the inflow turbulence characteristics. The effects of the inflow turbulence “flavours” and disc thrust coefficient on the virtual origin, which appears to be the leading parameter for wind turbine wake modelling, were finally examined. We emphasised the importance of carefully verifying the prerequisites of self-similarity and decaying turbulence within the Townsend-George theory before applying scaling laws to derive meaningful conclusions regarding the impacts of $C_T$ and the FST on the virtual origin. After ensuring these two conditions are met within the wakes, we observed that both turbulence intensity and integral length scale decrease the virtual origin, except for the wakes generated by D1, where the effects are opposite. Furthermore, we noted an increase in the virtual origin with increasing thrust coefficient

Since porous bodies were used as wake-generating objects in the present study, the distinct effects of thrust coefficient and porosity cannot be distinguished. The reversed effects of FST on the wake growth rate and entrainment rate observed with disc D1 may thus be attributed to its very high porosity or its very low thrust coefficient. For the specific set of discs used, we found this change at a critical porosity value of $\beta=0.6$, which equally corresponds to a critically low value of the thrust coefficient of $C_T=0.19$. Hence, determining whether the global porosity of a porous body, the distribution of porosity, or the thrust coefficient is the most important parameter in determining the wake dynamics remains an interesting open question. Moreover, further work should be carried out to determine if a similar change in how FST affects the wake dynamics and entrainment behaviour is equally observed with other types of porous bodies, such as wind turbines, for a critical thrust coefficient or porosity. Many experiments with wind turbines have been conducted at relatively high thrust coefficients, often near the optimal operating point. However, during their operational lifetime, wind turbines can also generate low thrust coefficients, typically in high wind speed conditions. Hence, exploring whether the effects of FST on wind turbine wakes remain consistent across their entire range of thrust coefficient and porosity, or if a similar singular behaviour is observed at low thrust coefficient/high porosity, could be of particular interest for wake modelling. In addition, determining to what extent the thrust coefficient alone is sufficient as a wind turbine parameter input in wake models, as well as how universally the porosity of a body affects entrainment mechanisms, needs further research.


To summarise, freestream turbulence has a clear effect on wake development downstream of porous bodies; both intensity and integral length scale are important, but it affects different wakes differently due to the importance of disc porosity and thrust coefficient. Discs with the lowest porosity behave in the same fashion as solid objects, both in terms of entrainment behavior and non-equilibrium scaling laws. As porosity increases, these “solid object effects" gradually diminish, yet similar trends persist, such as the suppression of entrainment in the far wake and the decrease in wake growth rate in the presence of FST. However, above a critical porosity, achieved in the present study with D1, the impact of FST on entrainment and wake growth is reversed. The very low velocity deficit and turbulence intensity produced by this highly porous disc, in comparison to the background turbulent characteristics, may contribute to this change in entrainment behaviour. Further investigation of this non-linear effects of the FST depending on the disc porosity and thrust coefficient could be addressed in future research.  Ultimately, we observed that the thrust coefficient and integral length had a strong effect on the wake growth rate as well as on the virtual origin; hence, implementing these findings into wind turbine wake models could be of great interest.

\begin{acknowledgments}

\textbf{Funding.} The authors gratefully acknowledge the Engineering and Physical Sciences Research Council (EPSRC) for funding this work through grant no. EP/V006436/1.

\textbf{Rights assertion statement.} For the purpose of open access, the authors have applied a Creative Commons Attribution (CC BY) licence to any Author Accepted Manuscript (AAM) version arising

\textbf{Declaration of interests.} The authors report no conflict of interest.

\end{acknowledgments}

\appendix

\section{Horizontal profiles of normalised mean velocity deficit and turbulence intensity \label{AppendixA}}

\begin{sidewaysfigure}
    \centering
    \includegraphics[width=\textheight,height=\textwidth,keepaspectratio]{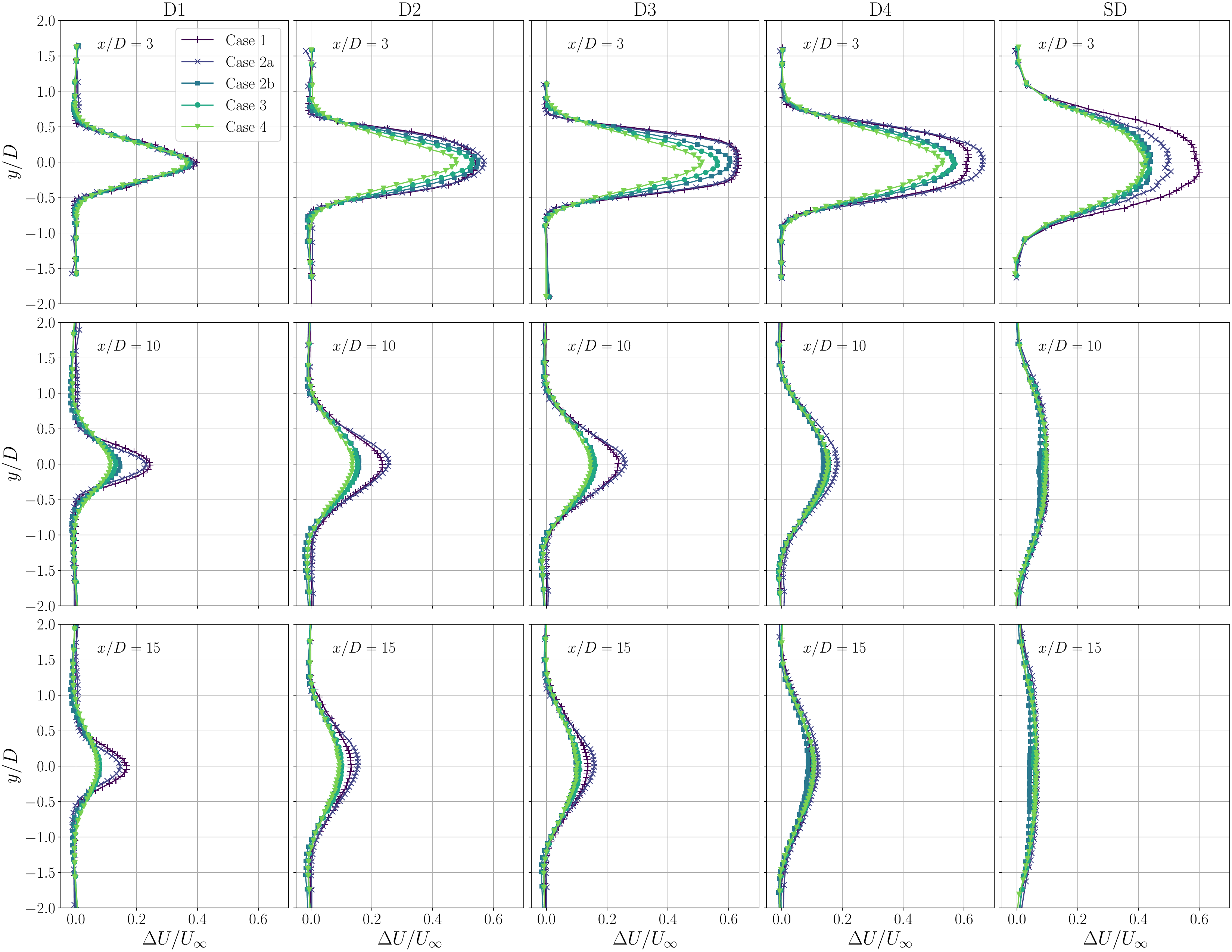}
    \caption{Horizontal profiles of normalised mean velocity deficit $\Delta U/U_{\infty}$ at 3 streamwise positions downstream of the discs and for the 5 different incoming turbulence “flavours”.}
    \label{Fig_AppendixA}
\end{sidewaysfigure}

\begin{sidewaysfigure}
    \centering
    \includegraphics[width=\textheight,height=\textwidth,keepaspectratio]{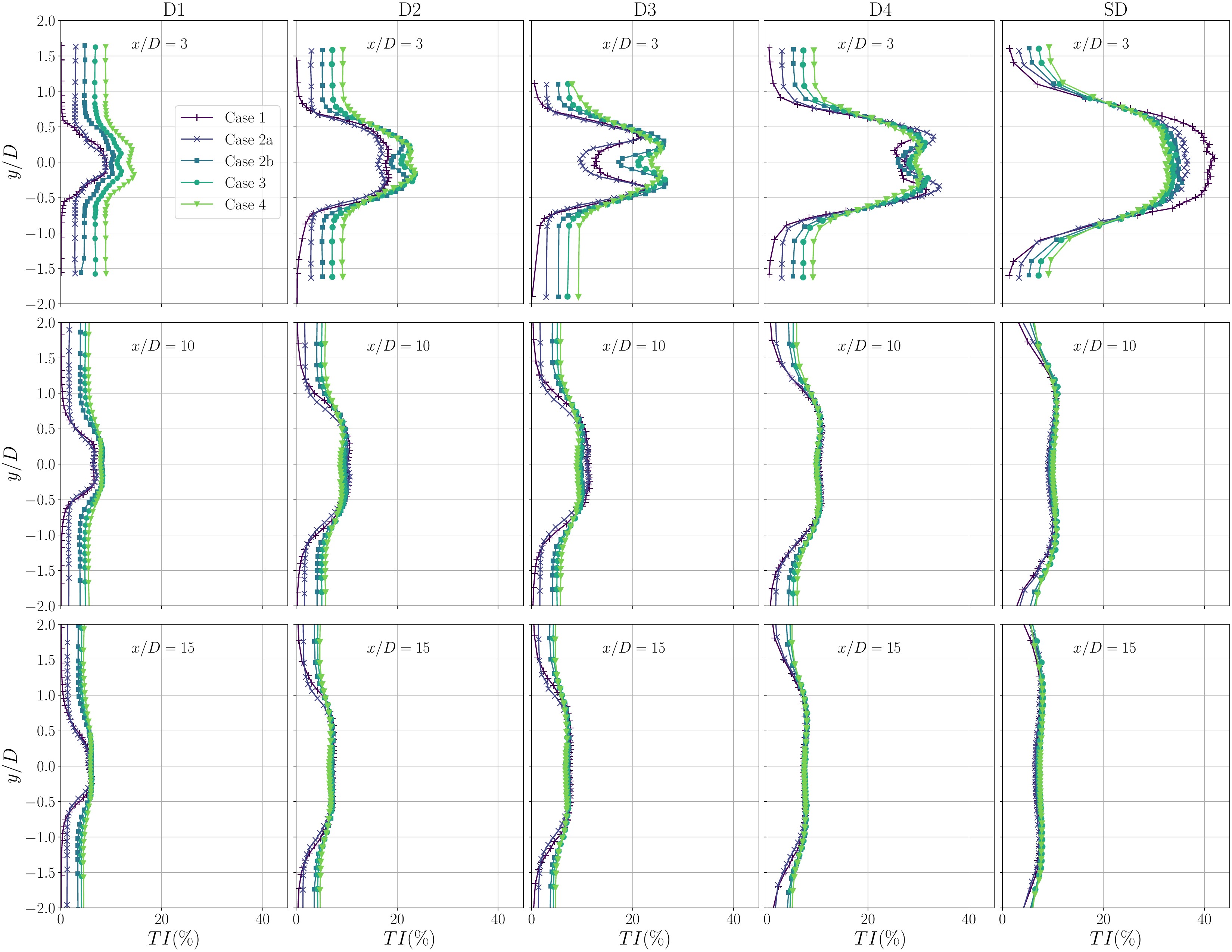}
    \caption{Horizontal profiles of turbulence intensity $TI$ 3 streamwise positions downstream of the discs and for the 5 different incoming turbulence “flavours”.}
    \label{Fig_AppendixB}
\end{sidewaysfigure}

\newpage


\bibliography{apssamp}

\end{document}